\documentclass[graybox]{svmult}

\usepackage{mathptmx}       
\usepackage{helvet}         
\usepackage{courier}        
\usepackage{type1cm}        
\usepackage{makeidx}         
\usepackage{graphicx}        
\usepackage{multicol}        
\usepackage[bottom]{footmisc}
\newcommand \bea{\begin{eqnarray}}
\newcommand \eea{\end{eqnarray}}
\newcommand \beq{\begin{eqnarray}}
\newcommand \eeq{\end{eqnarray}}
\newcommand \ga{\raisebox{-.5ex}{$\stackrel{>}{\sim}$}}
\newcommand \la{\raisebox{-.5ex}{$\stackrel{<}{\sim}$}}
\newcommand{\av}[1]{\langle{#1}\rangle}

\makeindex             

\begin{document}

\title*{Crossovers in Unitary Fermi Systems}
\author{Henning Heiselberg}
\authorrunning{H. Heiselberg} 
\institute{Henning Heiselberg \at Applied Research, DALO, Lautrupbjerg 1-5, DK-2750 Ballerup,
Denmark, \email{hh@mil.dk}}
\maketitle

\abstract{Universality and crossover is described for attractive and
  repulsive interactions where, respectively, the BCS-BEC
  crossover takes place and a ferromagnetic phase transition is
  claimed.  Crossovers are also described for optical lattices and
  multicomponent systems.  The crossovers, universal parameters and
  phase transitions are described within the Leggett and NSR models
  and calculated in detail within the Jastrow-Slater
  approximation. The physics of ultracold Fermi atoms is applied to
  neutron, nuclear and quark matter, nuclei and electrons in solids
  whenever possible.  Specifically, the differences between optical
  lattices and cuprates is discussed w.r.t. antiferromagnetic, d-wave
  superfluid phases and phase separation.}

\section{Introduction and the Bertsch problem}
\label{sec:1}

A decade ago at the 10th Manybody Conference G. Bertsch posed the problem 
(see \cite{MBX} for full text):
\begin{svgraybox}
How does a system of Fermi particles with infinite s-wave 
scattering length but vanishing interaction range behave?
\end{svgraybox}
This seemingly innocent question turned out to be rather fundamental and
triggered an explosion of interest
and number of papers on physics now referred to as 
\textit{universal physics, the unitarity limit, BCS-BEC crossover}, and strongly
  interacting and correlated systems in general.  His question was
partly motivated from nuclear physics and the physics of neutron stars
where dilute gases of neutrons exist in the inner crust. The
neutron-neutron $^1S_0$ scattering length $a\simeq -18.5fm$ is
long (and negative) compared to the order of magnitude
smaller interaction range $R\sim 1$fm, because
two neutrons almost have a bound state. 
Admittably, Bertsch was also
motivated by expectations that cold gases of Fermi atoms might be
created like the BEC a few years earlier, and that it might be
possible to tune interactions near Feshbach resonances between two hyperfine
states in order to
make the scattering length truly go to $\pm\infty$. Both were rapidly
and successfully accomplished in a number of remarkable experiments 
where also the BCS-BEC crossover,
multicomponent systems, optical lattices, etc. have been studied.

The solution to the Bertsch problem for two-component systems is
as simple as it is fundamental
\cite{Baker,long}:
\begin{svgraybox} Since $R\to0$ and $a\to\pm\infty$ both parameters must
vanish from the problem leaving only one remaining length scale: the
interparticle distance or equivalently the inverse of the Fermi
wavenumber $k_F^{-1}$. All thermodynamic quantities become \textit{universal}.
\end{svgraybox}
The proof is a dimensional argument which will be described in detail below.
As an example, the energy per particle is an \textit{universal} constant
times the energy of a non-interacting Fermi gas
$(3/5)E_F=3\hbar^2k_F^2/10m$.
The $a\to\pm\infty$
limit is referred to as the universal or \textit{unitarity} limit and the gas
as a \textit{unitary} gas. The name comes from scattering theory where
e.g. the s-wave scattering cross section at relative momentum $k$ is
limited by $\sigma_0\le 4\pi/k^2$ due to unitarity of the scattering
matrix. For weak interactions $\sigma_0=4\pi a^2$ which would diverge
like the energy per particle (see Eq. (\ref{Edilute})) - if extrapolated
to $a\to\pm\infty$. In the unitary gas the cross section and the
thermodynamic quantities are instead limited by unitarity and
universality respectively.

In the neutron gas $|a|$ is long but fixed and the unitarity limit is
defined as $|a|\gg r_0\gg R$, where $r_0=(3n/4\pi)^{-1/3}\sim
k_F^{-1}$ is the interparticle distance, $k_F$ the Fermi wavenumber at
density $n=k_F^3/3\pi^2$ for two components/spin states. Thus the
unitarity limit ranges from nuclear saturation density $n_{NM}\simeq
3/4\pi R^3\simeq 0.15$fm$^{-3}$, and more than three orders of
magnitude down in density to $\sim |a|^{-3}$.  In this region,
e.g. the energy per particle scales as an universal constant times
$E_F$. The unitarity limit became easily accessible experimentally as
it became possible to tune atomic interaction strengths near Feshbach
resonances \cite{Thomas}, where $a\to\pm\infty$ corresponding to
two-atom bound states at threshold.  The scattering length could also
be extended to positive values corresponding to bound molecular states
and for strongly bound states $a\to 0_+$ even a molecular BEC.

The Bertsch problem described above was originally intended for two component
system as spin 1/2 neutrons or ultracold atoms in two hyperfine states but
it is also relevant for a nucleon gas, since the neutron and a proton have the 
weakly bound state of deuterium
and therefore a large positive scattering length $a\simeq +20fm$.
Isospin symmetric
nuclear matter, however, has two spin and two isospin states i.e.
four components and does not have a unitarity limit. It is in fact unstable
at subnuclear matter densities where a liquid gas phase transition occurs.
This was successfully described in \cite{long} where estimates showed
that as Pauli blocking between same spins is effectively reduced as the number
of spin states increases, the Pauli pressure can no longer overcome the unitary
attraction. Only two component systems
are stable and provides a unitary gas. Three component systems are marginal 
and four and higher
component systems are unstable. Such multicomponent
systems are now studied, e.g. 
 $^6$Li with three hyperfine states \cite{Jochim}, $^{137}$Yb with six nuclear
spin states \cite{Kitagawa}, and heteronuclear mixtures of $^{40}$K and $^6$Li
\cite{Grimm}.
Multi-component systems have intriguing similarities with neutron, nuclear
and quark matter. In the latter color superconductivity between the
2 spin, 8 color and 2-3 flavor states may occur \cite{Alford}.

When the interactions are varied, e.g. near a Feshbach resonance, 
the thermodynamic quantities depend on the
dimensionless parameter $ak_F$. 
All the thermodynamic quantities become universal functions
of the crossover parameter $ak_F$ or equivalently $x=1/ak_F$ in the sense that they
do not depend on the system whether a gas of atoms, neutrons, or any other Fermi particle
as long as $R\ll |a|,r_0$.
The universality argument is intimately connected to the smooth approach to
the unitarity limit and crossover. At finite temperatures the universal thermodynamic
functions depend on $T/E_F$ as well.
A decade earlier pairing models \cite{Eagles,Leggett,NSR} had already described the 
crossover from BCS to the BEC limit and calculated pairing gaps, transition
temperatures and chemical potentials.

In these lecture notes universality, crossover and correlations will be
described. Not only the BCS-BEC crossover in uniform system but also
the repulsive ``ferromagnetic'' crossover, in multicomponent systems, traps and lattices.
Mostly Fermi atoms are discussed but applications to neutron, nuclear and quark
matter, nuclei and electrons in solids are made wherever possible.

\section{Universality and Crossover}
\label{sec:2}

\begin{figure}[t]
\includegraphics[scale=.75,angle=0]{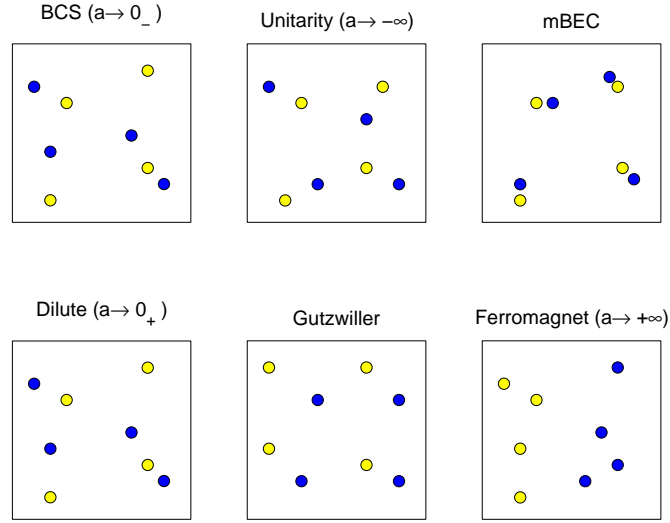}
\caption{Sketch of correlations in a two-component gas. Top row shows the gradual 
pairing of different components (spin states) with increasing attraction, 
i.e. the crossover from BCS via unitary gas to the molecular BEC. Bottow row shows 
the crossover for repulsive interactions from the dilute gas to either a hard sphere
(Gutzwiller) or a ferromagnetic phase (see text).}
\label{FigPhases}
\end{figure}

Universality is usually understood as the universal scaling in the unitarity
(Bertsch) limit where the interaction range ($R\to0$) and scattering length
($a\to\pm\infty$) are sufficiently short and large respectively that
the thermodynamic quantities are independent of either or any other
details of the interaction (such as effective ranges, etc.) and only
given by universal constants. 

When only the interaction range is
negligible ($R\to0$) but the scattering length varies, as it is possible in
ultracold atomic systems around Feshbach resonances, there is
another dimensionless parameter in the problem usually taken as $ak_F$ or
$x=1/ak_F$. The thermodynamic
quantities are thus universal functions of this 
\textit{crossover} parameter varying from weakly
interacting $x\to -\infty$ over the unitarity limit $x=0$ to the
strongly interacting molecular limit $x\to +\infty$.
In pairing models \cite{Leggett,NSR} these limits are named the BCS and BEC limits
respectively and the crossover as BCS-BEC crossover. 
When finite temperatures are considered the thermodynamic quantities
are universal functions of both $x$ and $T/E_F$ \cite{Ho}.

It should be emphasized that there are {\it several crossovers} (see Fig. (\ref{FigPhases}).
The most common one is that for increasingly attractive scattering lengths
also known from pairing models \cite{Leggett,NSR} and
referred to as the BCS-BEC crossover. Here the cold atoms start out in the weakly
attractive limit with BCS pairing i.e.  $x=1/ak_F\to -\infty$ and goes
over the unitarity limit $x=0$ to the
strongly attractive limit $x\to +\infty$, where a molecular BEC forms.
Another crossover starts out with weak repulsive interactions, i.e. a small positive
scattering length, and increasing it towards $ak_F\to+\infty$. A recent experiment
claims to observe a phase transition in a $^6$Li gas to a ferromagnetic state for
$ak_F\ga 2$ \cite{Jo}. These two crossover are different because the latter has
a node in the short range correlation function \cite{Cowell}. In principle there are a novel
crossover for each number of nodes, however, they become increasingly short lived due
to three-body losses. The different crossovers have different universal thermodynamic
functions, universal parameters and quite different superfluid and ferromagnetic phases.

We will in the following subsections describe universality and crossover
mainly for the BCS-BEC crossover and briefly for the ``ferromagnetic'' crossover. 
A number of thermodynamic functions are described which are
calculated within the Jastrow-Slater approximation for the many-body
wave function. 

\subsection{Universality in the unitarity limit}

The dimensional argument implies that any thermodynamic function, 
for example the energy per particle
in units of the Fermi energy, can at zero temperature
only depend on one dimensionless scale such as $x$
\begin{eqnarray} \label{Dense}
   \frac{E}{N} =   \frac{3}{5}E_{F}(1+\beta(x)) \,.
\end{eqnarray}
Here, the ratio between the interaction and kinetic energy
$\beta(x)$ is a ``universal'' many-body function
in sense that it and all other thermodynamic
functions are independent of the system, e.g. dilute neutron matter,
helium gases, atomic gases, etc.
Earlier definitions are $\eta=(3/5)(1+\beta)$
\cite{Carlson} and $\nu=(5/3)\beta$ \cite{long}.

The universality and crossover hypothesis implicitly assumes that
the transition to the unitarity limit and the crossover is smooth, i.e.
the derivatives of $\beta(x)$ are finite.
Taking the derivative with respect to density gives
\bea \label{dE}
  \frac{\partial E/N}{\partial n} =  \frac{3}{5}(1+\beta(x))
  \frac{\partial E_{G}}{\partial n}
    -\frac{E_{F}}{5n} x\beta'(x)
\eea
where $\beta'(x)=\partial\beta(x)/\partial x$.
We observe that derivatives of $\beta$ w.r.t. densities or $k_F$:
$k_F\partial/\partial k_F=-x\partial/\partial x$, 
always brings a factor $x$ or $1/a$. In the unitarity limit $x=0$, such terms
as the second term in Eq. (\ref{dE})
therefore vanishes and only the universal parameter $\beta(0)$ remains. 
By repeating the argument all density derivatives of
the energy such as chemical potential, pressure, first sound speed, compressibility, etc.
depend only on this one universal parameter $\beta(0)$.
Other thermodynamic variables than density derivatives of $E/N$ can, however, 
depend on higher derivatives $\beta^{(n)}(0)$ as we shall see below.

The first estimates based on Pade' approximants and Galitskii resummation
gave $\beta(0)\simeq-0.67$ \cite{Baker,long} in the unitary
limit whereas the Jastrow-Slater approximation yielded $\beta(0)\simeq-0.54$.
The Leggett pairing model, which will be
described in detail in a later section, predicts $\beta(0)\simeq-0.40$. 
Ground state energies have been calculated more accurately numerically
by Monte Carlo for systems with a finite number
of Fermions in a box. The earliest actually use the Jastrow-Slater
as trial wave functions for Green's function \cite{Carlson} 
and diffusion \cite{Casulleras} Monte Carlo.
Since pairing becomes important at crossover a more general
BCS wave function based on the Jastrow wave function was also employed.
Extrapolating to a large number of particles they obtained
$\beta(0)=-0.56\pm0.01$ and $\beta(0)=-0.58\pm0.01$ respectively
in the unitary limit.

Recent measurements \cite{Thomas,Ketterle,Bourdel,Regal,Chin} of $\beta(0)$
confirm the unitarity limit near Feshbach resonances.
Several experiments with trapped Fermi atoms have recently measured energies
in the strongly interacting or dense limit near Feshbach resonances.
The energy in the trap (excluding that from the harmonic oscillator potential)
is $E/N=(3/8)E_F\sqrt{1+\beta(0)}$ where $E_F=(3N)^{1/3}\hbar\omega$ 
is the Fermi energy in a trapped non-interacting gas.
The first measurements by the Duke group \cite{Thomas} measured 
the energy of $^6$Li Fermi atoms 
near a Feshbach resonance from expansion energies.
These early measurements were later  
corrected for thermal energies and find
$\beta(0)=-0.4\pm0.1$. \cite{Thomas,Bourdel} 
With the discovery of a molecular BEC the Innsbruck group
\cite{Chin} has been able to measure the size of the atomic
cloud, which scales with $(1+\beta)^{1/4}$, around the Feshbach
resonance at very low temperatures, and find $\beta(0)=-0.68\pm0.1$. 
Recent accurate measurements find $\beta(0)=-0.49\pm0.02$. \cite{Nascimbene} 

Other thermodynamic variables may in the unitarity limit
depend on other universal parameters.
For example, the number of closed channel molecules
is proportional to $\beta'(x)$ \cite{Hulet,Werner} and one obtains
$\beta'(0)\simeq -0.5$. From the slope of collective modes near the unitarity
limit $\beta'(0)\simeq -1.0$ \cite{Thomas,Bourdel}.
Generally, the universal function $\beta(x)$ is given by a Taylor expansion
in terms of an infinite number number of derivatives $\beta^{(n)}(0)$, which
all are universal parameters that determine all the universal thermodynamic functions.

Unlike neutrons, ultracold atoms can have many internal bound states
and corresponding Feshbach resonances
as the background magnetic field is increased.  Thus solutions are
multivalued for a given scattering length where the gas is in a metastable state.
For example, starting from a non-interacting gas ($a=0$) and increasing the
scattering length $a>0$ we also approach an unitarity limit $a\to+\infty$ \cite{Bosons}, 
which differs from that in the BCS-BEC crossover when the scattering length
is decreased $a<0$. 
This unitarity limit is similar to that for bosons as $a\to+\infty$ \cite{Cowell}, 
where the scattering length must be positive in order for the
system to be stable.
As will be described in detail the two-body wave function 
has a node and therefore the universal functions differs from those in the
BCS-BEC crossover. 
In principle a new universal limit and crossover exist for each
number of nodes, $n=1,2,3,...$, in the
two-body correlation function. These will, however, be increasingly unstable towards
three-body losses. The $n=1$ has been observed for bosons in the atomic-molecular
transition in a $^{85}$Rb BEC \cite{JILA} and is in agreement with the
predicted value from JS with $\kappa_1=2.80$. For fermions the predicted value
is calculated below within the Jastrow-Slater approximations
$\beta_1(0)\simeq2.93$ which is compatible with a recent experiment \cite{Jo}.
This value is about six times $|\beta(0)|$ demonstrating that these two
universal limits are very different.

\subsection{Thermodynamic crossover functions}

\begin{figure}[t]
\includegraphics[scale=.65,angle=-90]{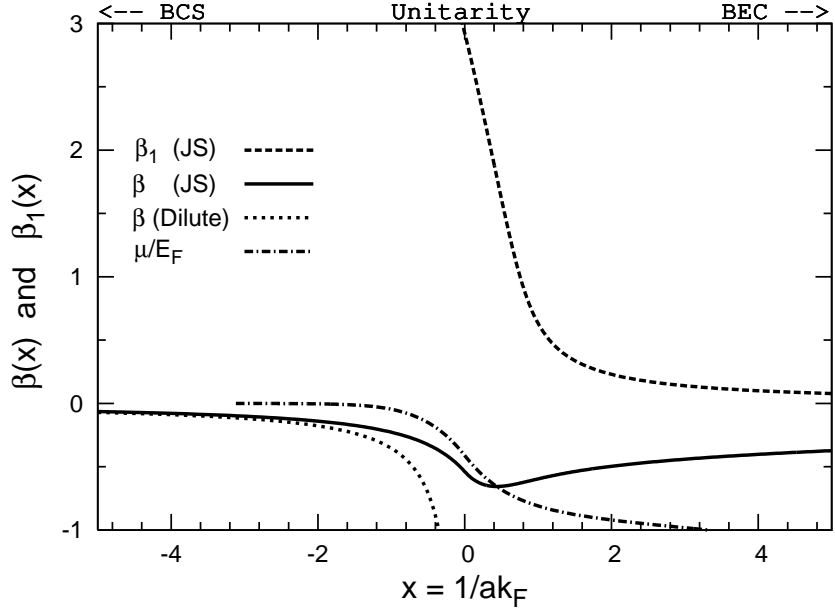}
\caption{The thermodynamic functions $\beta(x)$ and $\beta_1(x)$ within JS.
Also shown are the BCS limit of Eq. (\ref{beta}) and the chemical potential in
the Leggett crossover model (see text). On the BEC side ($x>0$) the molecular
energy been subtracted.}
\label{Figbeta}
\end{figure}

As argued above any thermodynamic function can only depend on one dimensionless
scale such as $x$ at zero temperature when the range of interaction is sufficiently small.
For example, the energy per particle
in units of the Fermi energy is given in Eq. (\ref{Dense}) 
in terms of the universal many-body crossover function
$\beta(x)=E_{int}/E_{kin}$. It is well known in
the dilute limit for Fermions \cite{LL}
\begin{eqnarray} \label{Edilute}
   \frac{E}{N} &=& \frac{3}{5} E_F+\frac{\pi\hbar^2}{m} an +...
\end{eqnarray}
to leading orders, whereas for molecular bosons
\begin{eqnarray} \label{Bdilute}
   \frac{E}{N} &=& -\frac{\hbar^2}{2ma^2} + \frac{\pi\hbar^2}{2m} a_Mn +... ,
\end{eqnarray}
where the molecular scattering length is $a_M\simeq 0.6a$ \cite{Petrov,Ketterle}.
Thus we obtain 
\bea \label{beta}
 \beta(x)=  \left\{\begin{array}{lll}
  10/(9\pi x)+4(11-2\ln2)/(21\pi^2x^2)+.. &,& x\to-\infty\\ 
  \beta(0)+\beta'(0)x+.. &,& x\simeq0
 \\ -5x^2/3-1+(a_M/a)5/(18\pi x)+..&,& x\to +\infty  
\end{array} \right\}.
\eea
In Fig. (\ref{Figbeta}) various model calculations of the universal function
$\beta(x)$ are shown.

The chemical potential is in terms of the universal function 
\bea
  \mu(x)=\left(\frac{\partial E}{\partial N}\right)_{V,S}
    =E_F\left(1+\beta-\frac{1}{5}x\beta' \right) \,.
\eea

In both the hydrodynamic limit and for a superfluid gas 
the first sound is given by the adiabatic sound speed
\bea
   c_S^2(x) = \frac{n}{m} \left(\frac{\partial \mu}{\partial n}\right)_{V,S}
      = \frac{1}{3}v_F^2 
  \left[ 1+\beta-\frac{3}{5}x\beta' +\frac{1}{10}x^2\beta''\right] \,,
\eea
where $v_F=\hbar k_F/m$.

The compressibility $\kappa=n^{-2}(\partial n/\partial\mu)$ is
related to first sound as 
\bea
\kappa^{-1}=n^{2}\left(\frac{\partial\mu}{\partial n}\right)_{V,S} = mnc_S^2 .
\eea

The pressure $P=n(\mu-E/N)=n^2dE/dV$ is
\bea \label{PF}
   P(x) =\frac{2}{5} E_F n[ 1+\beta-x\beta'/2] \,.
\eea

The polytropic index $P\propto n^{\gamma+1}$ is defined as
the logarithmic derivative of the pressure
\bea \label{gamma}
    \gamma(x) \equiv \frac{n}{P}\frac{dP}{dn} \, -1\, 
     =  \frac{\frac{2}{3}(1+\beta) -x\beta'/2
     +x^2\beta''/6}{1+\beta-x\beta'/2} \,.
\eea
$\gamma$ approaches 2/3 in both the dilute and 
unitarity limit whereas it approaches $\gamma=1$ in the molecular limit.
This index determines the frequency of collective modes in traps \cite{mode}.

We notice that all the above thermodynamic derivatives w.r.t. density only depend on
one universal parameter $\beta(0)$ in the unitarity limit $x=0$ as
argued above. Most thermodynamic quantities behave in the unitary gas
as in a free Fermi gas except for
the universal scaling factor $(1+\beta(0))$. This factor, however,
cancels in the polytropic index  and therefore
$\gamma=2/3$ in both the unitarity and BCS limit and the collective modes
are the same as has been verified experimentally \cite{Thomas}.

\subsection{Finite temperature}

At finite temperature the thermodynamic functions also depend on the
parameter $T/E_F$ (see, e.g., \cite{long,Ho,sound,Yu}).
Using $(\partial (E/N)/\partial T)_{V,N}=T (\partial s/\partial T)_{V,N}=c_V$,
where $s=S/N$ is the entropy and $c_V$ the specific heat per particle,
we obtain from Eq. (\ref{Dense})
\bea \label{betaT}
   \beta(x,T/E_F)= \beta(x,0)+\frac{5}{3}\int_0^T c_V(x,\frac{T'}{E_F}) \frac{dT'}{E_F} .
\eea

The entropy is at temperatures well below the superfluid transition temperature
$T\ll T_c$ given by phonon fluctuations
\bea
  s = \frac{2\pi^2}{45n} \left(\frac{T}{c_S}\right)^3 .
\eea
Inserting in eq. (\ref{betaT}) gives at low temperatures.
\bea \label{betaSF}
  \beta(x,T/E_F)= \beta(x,0)+5\frac{T}{E_F} s .
\eea
The temperature corrections scales as $\sim (T/T_F)^4$ and since $T\la T_c\simeq 0.19E_F$, 
the temperature dependence is almost flat as also observed in \cite{Nascimbene}.

In a Fermi liquid $s=(\pi^2m^*/2m)T/E_F$, where $m^*$ is the effective mass,
and thus 
\bea \label{betaFL}
 \beta(x,T/E_F)\simeq \beta(x,T_c/E_F)+\frac{5\pi^2}{6}\frac{m^*}{m}\frac{T^2-T_c^2}{E_F^2} ,
\eea
when $T_c<T\ll E_F$.
Recent accurate measurements find $\beta(0,0)=-0.49\pm0.01$ and $m^*/m=1.13\pm0.03$
\cite{Nascimbene}
in the unitarity limit. These allow us to extract the
Landau parameters at zero temperature
$F_0^s=(1+\beta(0,0))m^*/m-1\simeq -0.42$ and $F_1^s=3(m^*/m-1)\simeq 0.39$.
In the BCS limit $m^*/m=1+[8(7\ln2-1)/15\pi^2]a^2k_F^2$ and
$F_0=1+(10/9\pi)ak_F$.

As the temperature decreases towards the
critical temperature the change in $\beta(T)$ from a Fermi liquid Eq. (\ref{betaFL})
towards a superfluid Eq. (\ref{betaSF}) 
allowed determination of 
$T_c/E_F=0.19\pm0.02$ \cite{Nascimbene}.

At temperatures below $T<T_c$ the normal and superfluid components lead to
two (first and second) sound modes in the collisional limit.
In the BCS and BEC limit these are also referred to as the compressional and thermal
sound modes.
Their velocities $u_1$ and $u_2$ 
are given by the positive and negative solutions respectively of  
\cite{Zaremba,PS}
\bea \label{sss}
  u^2 = \frac{c_S^2+c_2^2}{2}\pm
       \sqrt{\left(\frac{c_S^2+c_2^2}{2}\right)^2-c_T^2c_2^2} \,.
\eea
The thermodynamic quantities entering are
the adiabatic $c_S^2=(\partial P/\partial n)_S$ and the
isothermal $c_T^2=(\partial P/\partial n)_T$ compressional sound speed squared.
The ``thermal'' sound wave $c_2^2=n_s s^2T/n_n c_V$ 
acts as a coupling or mixing term and is small in the two limits.
Here, $n=n_n+n_s$ is the total,  $n_n$ the normal and $n_s$ 
the superfluid density.
The difference between the adiabatic and isothermal
sound speed squared can also be expressed as
\bea
  c_S^2-c_T^2=\left(\frac{\partial s}{\partial n}\right)^2_T 
           \frac{n^2T}{c_V}  \, .
\eea
The mixing of the compressional and thermal
sound modes is particular interesting at crossover
where they mix and couple strongly. They undergo avoided crossing, i.e. the
compressional change smoothly into a thermal sound mode and visa versa
around the unitarity limit \cite{sound}.

\subsection{Jastrow-Slater approximation}

The Jastrow and Jastrow-Slater (JS) approximation
was among the earliest models applied to the unitarity limit and
crossover \cite{long}. 
It has the advantage that it provides analytical results for the universal
parameters that are easy to understand and has proven to be quite accurate
when compared to experiment.
It gives an ansatz for the strongly correlated wavefunctions which 
is also the starting point as trial wavefunctions in
Monte Carlo calculations \cite{Casulleras}. Finally, JS can be generalized to describe
both bosons and fermions with any number of spin states as well as other universal
limits.

The Jastrow and the JS approximation methods were developed for strongly interacting and correlated
Bose and Fermi fluids respectively such
as $^4He$, $^3He$ and nuclear matter in \cite{Vijay} and has
more recently been applied to kaon condensation \cite{kaon},
cold atomic Fermi \cite{Bosons} and Bose \cite{Cowell} gases. 
As explained in these references the JS wave function 
\begin{eqnarray}
\Psi_{JS}({\bf r}_1,...,{\bf r}_N)= 
\Phi_S\prod_{i,j'}\phi({\bf r}_i-{\bf r}_{j'}) \,,
\end{eqnarray} 
incorporates essential two-body correlations in the Jastrow function $\phi(r)$.
The antisymmetric Slater wave function $\Phi_S$ for free fermions
$\Phi_S$ insures that same spins are spatially anti-symmetric. 
The Jastrow wave function only applies to particles with different spins
(indicated by the primes). The pair
correlation function $\phi(r)$ can be determined variationally by
minimizing the expectation value of the energy, $E/N = \langle \Psi
|H| \Psi \rangle \ /\ \langle \Psi | \Psi \rangle$, which may be
calculated by Monte Carlo methods that are fairly well approximated by
including only two-body clusters. The basic idea of this method
is that at short distances $r \ll r_0$
the Jastrow function $\phi(r)$ obeys the
Schr\"odinger equation for a pair of particles
interacting through a potential $U(r)$ 
\bea \label{Schrodinger}
  \left[-\frac{\hbar^2}{m}\frac{d^2}{dr^2} +U(r)\right] r\phi(r) 
 =\epsilon_M\, r\phi(r)
   \, ,
\eea
where the eigenvalue energy of two atoms $\epsilon_M=2E_{int}/N$.
Many-body effects become important,
when $r$ is comparable to $r_0$, but are found to be small
in lowest order constrained variation 
(LOCV \cite{Vijay,long,kaon,Bosons}). Here the boundary conditions
that $\phi(r>d)$ is constant and $\phi^{\prime}(r=d)=0$ 
are imposed at the healing distance $d$, 
which is determined self consistently from number conservation
(see \cite{Bosons} for details)
\bea \label{numberspin}
  \frac{\nu-1}{\nu} n\int_0^d \frac{\phi^2(r)}{\phi^2(d)} 4\pi r^2dr = 1
\eea
Note that a given component only interact
and correlate with the $(\nu-1)$ other components which explains the prefactor.
In the dilute limit $\phi(r)\simeq 1$ and so $d=(\nu/(\nu-1))^{1/3}r_0$.
In the unitary limit $a\to \pm\infty$ 
the healing length now approaches
$d=r_0(2\nu/(\nu-1)3)^{1/3}=(3\pi/(\nu-1))^{1/3}k_F^{-1}$.
Generally $d\simeq r_0$.
The boundary condition at short distances is
given by the scattering length $(r\phi)'/r\phi=-1/a$ at $r=0$.

\subsubsection{Attractive crossover}

\begin{figure}[t]
\includegraphics[scale=.65,angle=-90]{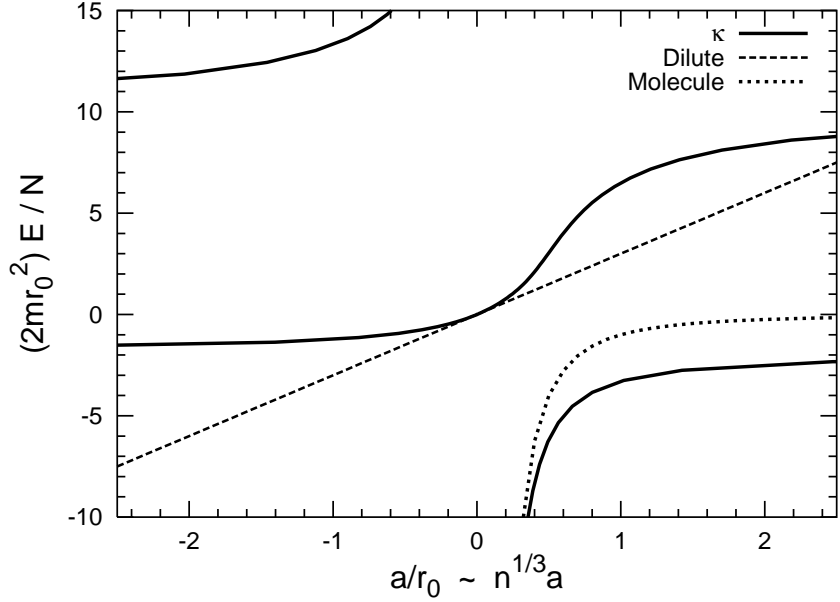}
\caption{$\kappa^2$ vs. scattering length in the JS model in units of interparticle distance
$r_0$.}
\label{Feshbach}
\end{figure}

\begin{figure}[t]
\includegraphics[scale=.65,angle=-90]{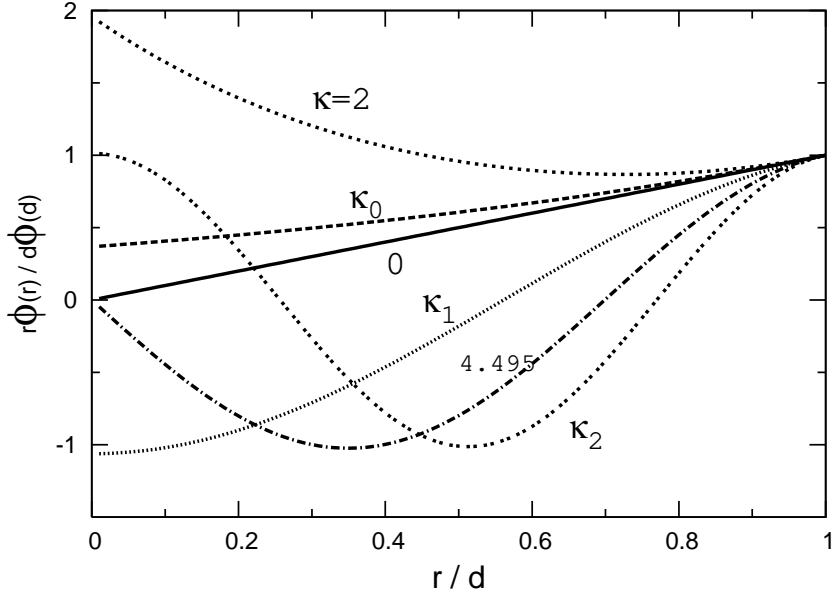}
\caption{The JS wave function $r\phi(r)/d\phi(d)$ in the dilute limit
  ($\kappa=0$), the attractive unitarity limit ($\kappa_0$), towards
  the mBEC ($\kappa=2$), the repulsive unitarity limit with one
  ($\kappa_1$) and two nodes ($\kappa_2$).}
\label{Figwf}
\end{figure}

For negative scattering lengths or negative energies $\epsilon_M<0$
the solution to the Schr\"odinger equation gives a wave-function
$r\phi(r)\propto\sinh[k(r-b)]$ for weak interactions which change to 
$r\phi(r)\propto\cosh[k(r-b)]$) for stronger interactions (see Fig. (\ref{Figwf})).
The boundary conditions and number conservation
determine the phase $kb$, the energy and the healing length $d$. 
For small scattering lengths $b=a$ whereas $b=0$ in the unitarity limit.
The interaction energy $E_{int}/N=-\hbar^2k^2/2m$ is given by \cite{A}
\bea \label{neg}
 \frac{a}{d} = \frac{\kappa^{-1}\tanh\kappa-1}{1-\kappa\tanh\kappa} \,,
\eea
with $\kappa=kd$. 
In the BCS limit Eq. (\ref{neg}) 
gives the correct interaction energy per particle, Eq. (\ref{Edilute}).
In contrast
the negative energy solution
to Eq. (\ref{neg}) reduces in the unitarity limit to
$\kappa\tanh\kappa=1$ with solution $\kappa_0=1.1997...$
As the scattering length cross over from $-\infty$ to $+\infty$ the negative
energy state is analytically continued towards the molecular bound state
with $E/N=-\hbar^2/2ma^2$ as $a\to +0$.

In addition to the interaction energy $E_{int}=\kappa^2/2md^2$ as calculated above
a kinetic energy $(3/5)E_F$ appears due to the Slater
ground state. The total energy becomes
\begin{eqnarray} \label{LOCV-F}
  \frac{E}{N}&=&\frac{3}{5}E_F - \frac{\hbar^2\kappa^2}{2md^2} \,.
\end{eqnarray}
From the definition Eq. (\ref{beta}) we obtain the universal function
\bea
  \beta = -\frac{5}{3} \left(\frac{\kappa}{dk_F}\right)^2 .
\eea
In the unitarity limit $\kappa=\kappa_0=1.1997$ and $dk_F=(3\pi/(\nu-1))^{1/3}$ 
and the universal parameter is for a general number of spin states
\begin{eqnarray} \label{betaJS}
    \beta(0) &=&  -\frac{5}{3} 
    \left(\frac{\nu-1}{3\pi}\right)^{2/3} \kappa_0^2   \,.   
\end{eqnarray}
Thus in a two-component system $\beta(0)=-0.54$ which lies between
Monte Carlo results $\beta(0)\simeq-0.56$ \cite{Casulleras} and recent experimental data
$\beta(0)\simeq-0.49$ \cite{Nascimbene}.

The universal function $\beta(x)$ is shown in Fig. (\ref{Figbeta}) for
the JS approximation.  Also shown is the chemical potential within the
Leggett crossover model which will be discussed below in connection
with pairing.  The JS model includes self energies and is therefore a
better approximation on the BCS side ($x<0$). On the BEC side ($x>0$)
both models approach the molecular energy (subtracted in
Fig.  (\ref{Figbeta})) but to next orders both model overestimates the
energy: the JS model by the Slater energy and the Leggett model by
overestimating the molecular scattering length $a_M$ by a factor
$\sim3$.

The dependence on the scattering length is given by
Eq. (\ref{neg}). By taking the derivative on both sides w.r.t. $x$ we obtain
\bea \label{dbeta}
  \beta'(0) = -\frac{10}{3}\frac{1-\kappa_0^{-2}}{(3\pi)^{1/3}} \simeq -0.48 \, .
\eea
This JS prediction is somewhat lower than that of Monte Carlo: $\beta'(0)=-1.0\pm0.1$
\cite{Carlson,Casulleras}. The slopes of the axial and longitudinal
collective frequencies of trapped unitary Fermi gases are directly proportional
to $\beta'(0)$ and measurements also give $\beta'(0)\simeq-1.0$ \cite{Thomas,Bourdel}.
Analysis of the number of closed channel molecules indicates
$\beta'(0)\simeq-0.5$ \cite{Werner,Hulet}.
The Leggett pairing model $\beta'(0)\simeq-1.0$.

The universal parameter $\beta'(0)$ will be related to short range correlations
in sec. (\ref{SRcorr}).

\subsubsection{Repulsive crossover}

Starting from a positive scattering length the and positive interaction energy 
the solution to the Schr\"odinger equation gives a wave-function
$r\phi(r)\propto\sin[k(r-b)]$ (see Fig. (\ref{Figwf})). The boundary conditions and number conservation
gives \cite{Cowell}
\bea \label{pos}
 \frac{a}{d} = \frac{\kappa^{-1}\tan\kappa-1}{1+\kappa\tan\kappa} \,.
\eea
As in the BCS limit Eq. (\ref{pos}) 
gives the correct interaction energy per particle in the dilute limit, Eq. (\ref{Edilute}).
In the unitarity limit $a\to +\infty$, the positive energy
solution reduces to $\kappa\tan\kappa=-1$ with multiple solutions 
$\kappa_1=2.798386..$, $\kappa_2=6.1212..$, etc., 
and asymptotically $\kappa_n=n\pi$ for integer $n$. Generally, $n=0,1,2,..$ is the number
of nodes in the JS wavefunction and each determines a new universal limit with
universal parameters depending on the number of nodes. The phase in the wave function
is $kb=\pi(n-1/2)$ whenever the unitarity limit of  $n$ nodes is encountered.

The resulting energy is
\begin{eqnarray} \label{LOCV+}
  \frac{E}{N}&=&\frac{3}{5}E_F + \frac{\hbar^2\kappa^2}{2md^2} \,,
\end{eqnarray}
and we obtain from Eq. (\ref{Dense}) 
\bea
  \beta(x) = \frac{5}{3} \left(\frac{\kappa}{dk_F}\right)^2 .
\eea
Because Eq. (\ref{pos}) have a string of solutions for given $a$, 
$\kappa$ and $\beta$ are multivalued functions for a given scattering length or $x$,
which we distinguish by an index
$\beta_n$ referring to the number of nodes $n=0,1,2,...$. The attractive case discussed
above is the $n=0$ case, i.e. universal function in the BCS-BEC crossover is
also $\beta=\beta_0$. $n=1$ is the first weakly repulsive crossover where a
ferromagnetic phase transition may occur and is also shown in Figs. (\ref{Figbeta}) and
(\ref{Feshbach}).

The universal parameter in this repulsive unitarity limit is
\bea
 \label{beta+}
    \beta_1(x=0) &=&  +\frac{5}{3} 
    \left(\frac{\nu-1}{3\pi}\right)^{2/3} \kappa_1^2\simeq 2.93(\nu-1)^{2/3}   \,.  
\eea
The index denotes that this universal parameter is for the wave function with
one node. It has recently been measured in 
a $^6$Li gas in two spin states \cite{Jo}. 
The chemical potential in the optical trap almost doubles
going from the non-interacting to the unitarity limit. As shown below it scales
as $\mu\propto\sqrt{1+\beta_1(0)}$ giving $\beta_1(0)\sim 3$ compatible with
JS. However, the experiments also indicate a
phase transition in repulsive unitarity limit from a paramagnetic to a ferromagnetic phase.
This phase transition is supported by the JS models as will be shown in sec. (\ref{ferro}).

As for the attractive crossover we calculate within JS the derivative from Eq. (\ref{pos})
\bea \label{dbeta+}
  \beta'_1(0) = -\frac{10}{3}\frac{(1-\kappa_1^{-2})}{(3\pi)^{1/3}} \simeq- 1.38 \, .
\eea

\subsection{Short range correlations}
\label{SRcorr}

The short range two-body correlations are connected to thermodynamic
quantities of the system \cite{Fisher}. We shall here derive one
important example at zero temperature and use the Jastrow-Slater
results above to actually calculate a new universal number.
We refer to \cite{Tan,Yu} for finite temperature extensions.

Following \cite{Yu} we scale the two-body potential by a factor $\lambda$
and obtain from Feynmann-Hellmann
\bea \label{FH}
  \frac{\partial E/V}{\partial\lambda} = \frac{1}{V}
  \langle\frac{\partial H}{\partial\lambda}\rangle 
 = \int d^3r U(r) 
 \langle \psi_1^\dagger({\bf r})\psi_2^\dagger(0)\psi_2(0)\psi_1({\bf r})\rangle_\lambda .
\eea
Here the correlation function at short ranges is
\bea \label{Corr}
  \langle \psi_1^\dagger({\bf r})\psi_2^\dagger(0)\psi_2(0)\psi_1({\bf r})\rangle_\lambda
 = \phi^2_\lambda(r) = C\left(\frac{1}{r}-\frac{1}{a_\lambda}\right)^2 ,
\eea
where $\phi_\lambda$ is the solution to the
two-body Schr\"odinger equation (\ref{Schrodinger}) 
but with potential $\lambda U(r)$.
$C$ is a normalization factor such that
$(1/V)\int d^3 r \phi_\lambda^2(r)=(n/2)^2$, and
will later  be interpreted as the correlation strength. 

Taking the derivative of the Schr\"odinger equation 
w.r.t. $\lambda$, multiplying by $r\phi_\lambda$, and integrating over
$r$ we obtain
\bea \label{C}
  \int_0^{r_c} dr U(r)(r\phi_\lambda(r))^2 = -\frac{C}{m}  
  \frac{\partial a^{-1}}{\partial\lambda} ,
\eea
where $r_c$ is any range between $R\ll r_c\ll r_0$.
Combining Eqs. (\ref{FH}) and (\ref{C}) gives
\bea
  \frac{\partial E/V}{\partial a^{-1}} = -\frac{4\pi C}{m} .
\eea
From Eq. (\ref{Dense}) we can relate the correlation strength to the
derivative of the universal functions $\beta_n(x)$ as
\bea
   C_n =-\frac{k_F^4}{40\pi^3} \beta'_n(x)  ,
\eea
where the node index $n$ distinguishes the various universal functions as discussed above.
In the unitarity limits we can determine $C_n(x=0)$ by inserting
$\beta'_0(0)=-0.54$ or $\beta'_1(0)=-1.38$
from JS Eqs. (\ref{dbeta}) and (\ref{dbeta+}), etc.
$C_n(x)$ can also be
obtained directly from the JS wave-function, normalization and Eq. (\ref{Corr}).

The Fourier transform of 
the wave-function at large momenta is determined by the short range 
correlations  $\phi(k)=\int d^3r e^{i{\bf k\cdot r}}\phi(r)\simeq 4\pi\sqrt{C}/k^2$. 
Therefore the momentum distribution for a spin state has the tail
\bea
   n_\sigma(k) = \phi^2(k)= (4\pi)^2C/k^4 ,
\eea
at large momenta as is well known for Fermi liquids \cite{LL}.
The correlation strength is directly measured in 
the number of closed channel molecules \cite{Hulet} 
\bea
  N_M= \frac{\partial E}{\partial a^{-1}}\frac{\partial a^{-1}}{\partial B}
   \frac{1}{\mu_M} ,
\eea
where $\mu_M$ is twice the magnetic moment of the atom.
The scattering length depends on the applied magnetic field 
as $a\simeq a_{bg}[1-\Delta B/(B-B_0)]$ near a Feshbach resonance at $B_0$.
Experiments find for the BCS-BEC crossover \cite{Hulet} 
$\beta'_0(0)\simeq -0.5$ which agrees well with that of JS and is
compatible with model estimates \cite{Werner}.

\subsection{Instability in multicomponent systems}
\label{instability}
The original Bersch problem was intended for two-component systems
as neutron matter or spin-balanced ultracold atoms in two hyperfine states.
Early extrapolations of the dilute energy of Eq. (\ref{Edilute}) to $a\to-\infty$
led to the belief that the unitarity limit was unstable towards collapse
\cite{Houbiers} as a nuclear gas
whereas in fact the two-component system is the only stable unitary gas.
The predicted scaling for the neutron gas at subnuclear
density as given by the unitary gas of Eq. (\ref{Dense}) triggered the
memory of Vijay Pandharipande\cite{Nordita}, who had the impression
that Bethe and Brueckner were aware of this scaling when they looked
at low density neutron matter in neutron star context back in the
'50s \cite{BBG,Vijay}.\footnote{However, no written reference has been found 
in the works of Bethe, Brueckner or Pandharipande so far.}
Calculations by Carlson et al. \cite{Carlson}  have later confirmed
this scaling in neutron gases.  

Multicomponent systems are now studied, e.g. $^6$Li with three
hyperfine states \cite{Jochim}, $^{137}$Yb with six nuclear spin
states \cite{Kitagawa}, and heteronuclear mixtures of $^{40}$K and
$^6$Li \cite{Grimm}. Such multi-component systems have intriguing
similarities with neutron, nuclear and quark matter where
color superconductivity between the 2 spin, 8 color and 2-3
flavor states may occur \cite{Alford}.

For the gas to be stable towards collapse in the unitarity limit
the energy must be positive,
i.e., $1+\beta(0)>0$. In the JS approximation we obtain in the unitary
limit from Eq. (\ref{betaJS}) $\beta(0)=0,-0.54, -0.85,-1.12,....$,
for $\nu=1,2,3,4,...$ spin states respectively.  Therefore the JS
approximation predicts that up to $\nu\le3$ spin states are stable in
the unitary limit whereas in the Galitskii approximation only
$\nu=1,2$ are stable \cite{long}. Pauli blocking is effectively reduced in many
component system and can only stabilize one, two and perhaps three
component Fermi systems in the unitary limit. 
The stability of two spin states towards collapse has been confirmed
for a $^{6}$Li and $^{40}$K gases near Feshbach resonances. 
 The marginal case
$\nu=3$ has been studied with $^6$Li atoms, which have three spin
states with broad and close lying Feshbach resonances, and
although the loss rate of atoms is large near Feshbach and Efimov
resonances \cite{Jochim} the gas is sufficiently long lived for measurements
and does not collapse.
It has long been known that neutron star matter \cite{NS} with two spin
states likewise has positive energy at all densities whereas for
symmetric nuclear matter with two spin and two isospin states, i.e.
$\nu=4$, the energy per particle is negative. Nuclear matter is
therefore unstable towards collapse and subsequent implosion, spinodal
decomposition and fragmentation at subnuclear densities\cite{HPR}.
Above nuclear saturation densities, $k_FR\ga1$, short range repulsion
stabilizes matter up to maximum masses of neutron stars $\sim
2.2M_\odot$, where gravitation makes such heavy neutron stars unstable
towards collapse \cite{NS}.
The conjecture is therefore that the $\nu>3$ Fermi systems
are unstable and non-universal in the unitarity limit $a\to -\infty$.
Bose atoms corresponds to $\nu=\infty$, since no Pauli blocking applies,
and are unstable for negative scattering lengths.

The $\nu=3$ system is relevant for several reasons.  Traps with
ultracold $^6$Li atoms with three hyperfine states are sufficiently
stable and long lived \cite{Jochim} to be studied in detail. The three
Feshbach resonances are, however, separate in magnetic field so that
the unitarity limit is not simultaneous for the three components.
The three body system has interesting Efimov states which are  
non-universal \cite{Efimov,Braaten}, i.e. besides the scattering length
the system depends on an additional
potential parameter such as the effective range. This
non-universality persists for three bosons confined in a trap \cite{Jonsell}
and non-universality is therefore expected for a gas with three state Fermi atoms
as experiments for the $^6$Li system also indicates \cite{Jochim}.
Therefore three component systems do not have an universal limit
and are only marginally stable, i.e. stable towards collapse
but suffer three-body losses.

In spin polarized systems or systems with different densities of the
spin components the stability conditions depends on the various
component densities.  Also the system may undergo phase separation
into a more symmetric phase and an asymmetric phase as, e.g., nuclei
and a neutron gas in the inner crust of neutron stars. Similarly, spin
polarized atoms at low temperature in traps may for strong attractive
interactions separate into a paired spin balanced phase in the centre
with a mantle of excess spin atoms.  Likewise phase separation of
strongly repulsive fermions may separate into domains of ferromagnetic
phases in the centre with a paramagnetic mantle around.

\subsection{Repulsive interactions and itinerant Ferromagnetism}
\label{ferro}

For a small positive scattering length the Fermi gas is a paramagnet (PM).
For stronger repulsion Stoner \cite{Stoner} predicted a phase
transition to a ferromagnet (FM)
which a recent experiment claim to have observed \cite{Jo}.
Stoner's argument was based on the dilute equation of state of Eq. (\ref{Edilute}) which
generally for a spin polarized two-component system of total density $n=n_\downarrow+n_\uparrow$
and polarization $\eta=(n_\downarrow-n_\uparrow)/n$ is
\bea
 E/N &=&\frac{3}{10} E_F\left[ (1+\eta)^{5/3}+(1-\eta)^{5/3}
     +\frac{20}{9\pi}(1+\eta)(1-\eta)k_Fa \right] .
\eea
Expanding for small polarization gives an equation of the Ginzburg-Landau type
\bea \label{GL}
   E/NE_F &\simeq&  \frac{3}{5}+\frac{2}{3\pi}ak_F+
        \frac{1}{3}(1-\frac{2}{\pi}ak_F)\eta^2 +3^{-4}\eta^4 +{\cal O}(\eta^6).
\eea
It predicts a second order phase transition at $ak_F=\pi/2$ from a PM to a FM
with polarization $\eta=\pm\sqrt{27(2ak_F-1/\pi)}$. Due to the small fourth order
coefficient it quickly leads to a 
locally fully polarized system $\eta=\pm1$.

Unfortunately the predicted transition occurs close to the unitarity
limit where the dilute equation of state is not valid. Higher orders may be important
as exemplified by including the next order correction of order $a^2$. It
changes the transition from second to first order \cite{Conduit} at low
temperatures up to a tri-critical point at $T_c\simeq 0.2T_F$, where the
transition becomes second order again. However, the dilute expansion remains invalid
in the unitarity limit.

The Jastrow-Slater approximation extends to the unitarity limit also for positive
scattering lengths as discussed above. 
Number conservation of the various spin densities is automatically included in
the healing length, see e.g. 
Eq. (\ref{numberspin}). As result the energy is
\bea 
\frac{E}{N} &=&\frac{3}{10}E_F \left[ (1+\eta)^{5/3}+(1-\eta)^{5/3} 
     +\beta_1 (1+\eta)(1-\eta)^{2/3}  +\beta_1(1+\eta)^{2/3}(1-\eta)\right] 
 \nonumber\\
 && +{\cal O}(\eta^6).
\eea
$\beta_1(x,\eta)$ is now also a function of polarization but
for simplicity we shall ignore the dependency this dependency as we expect it to be minor.
Expanding for small polarization we find
\bea
   E/NE_F &\simeq&  \frac{3}{5}(1+\beta_1) +
        \frac{1}{3}(1-\frac{7}{5}\beta_1)  \eta^2 +3^{-4}(1-\beta_1)\eta^4 \,.
\eea
Truncating to order $\eta^4$ would (erroneously) predict a second order phase transition from a PM to a FM
at $\beta_1(x)=5/7\simeq 0.71$.
The fourth order term is even smaller than the dilute prediction
of Eq. (\ref{GL}), and it is
therefore necessary to include higher orders.
By equating the energy of the unpolarized gas, $\sim(1+\beta_1)$ with that of
a fully polarized gas, $\sim 2^{2/3}$, we find a first order transition
at $\beta_1=2^{2/3}-1\simeq0.59$, since this value is smaller than 5/7.
In view of the approximations made in this model calculation and the proximity
of the two $\beta_1$ values for the first and second order transitions, we can 
not reliably determine whether the order of the PM-FM transition is first or second.
However, since the repulsive interaction energy in the unitarity limit
$\beta_1(0)\simeq \kappa_1^2/(3\pi)^{2/3}\simeq2.93$.
is much larger than the critical value $\beta_1\simeq0.59$ 
we can safely conclude that the transition to a FM does take place
at a value corresponding to  $ak_F\simeq 0.85$.

In a recent experiment the transition is observed around $ak_F^0\simeq
2.2$ at temperatures $T/T_F=0.12$ and $ak_F^0\simeq 4.2$ at
$T/T_F=0.22$ \cite{Jo}. Presumably, the critical value for $ak_F^0$ is
smaller at zero temperature. Also the Fermi wavenumber $k_F^0$ is the
central value for a non-interacting gas which is larger than the
average value over the trap of the gas that is further expanded due to
repulsive interactions. More experiments will determine
$\beta_1(x)$, a possible ferromagnetic phase and critical value for $ak_F$.

When the number of atoms in the two spin states is balanced the ferromagnetic
domains of $\eta=\pm1$ coexist.
Their domain sizes may be to small to observe in present
experiments \cite{Jo}. The densities of the two components will, however,
have interesting distributions for unbalanced two-component systems
in traps, where the minority component will be suppressed in the centre and
both phase separation and ferromagnetism can occur.
In three component systems,
when there are more than one Feshbach resonance as in
$^6$Li, with Feshbach magnetic field 
such that two resonances $a_{12}$ and $a_{13}$ are large but $a_{23}$ small,
the atoms will separate between a FM phase of 1 and
a mixed FM phase of 2+3 with different densities.

It should be emphasized that for positive scattering lengths
the wave function and thus the correlations function
between fermions of unlike spin and bosons $\chi=r\phi\sim\sin(kr-b)$ has a node
somewhere within the interparticle distance $[0;r_0]$ 
(see \cite{Cowell} and Fig. (\ref{Figwf})). 
It does not vanish
as $r\to0$ as does the wave function for a short range repulsive potential as in 
hard sphere scattering, where $a\simeq R$.
Therefore the Gutzwiller approximation discussed in \cite{Zhai} applies to
hard sphere gases, strongly correlated nuclear fluids and liquid helium but not
to the repulsive unitarity limit of ultracold gases (see Fig. (\ref{FigPhases}).

\section{Pairing in uniform systems and the BCS-BEC Crossover}

Experiments on superfluid Fermi gases have recenty confirmed the
BCS to BEC crossover models \cite{Eagles,Leggett,NSR,Strinati,Bloch} 
for pairing that was developed after the invention of BCS theory.
The development of BCS-BEC crossover will be described historically with
increasing level of complexity.

\subsection{BCS limit}

Bardeen, Cooper and Schriffer (BCS) first wrote down the famous gap
equation \cite{BCS} for an attractive two-body interaction $U(r)<0$
\bea \label{BCS}
\Delta_{\bf p} = -\frac{1}{V}\sum_{\bf p'} U({\bf p',p}) 
           \frac{1-2f(E_{\bf p'})}{2E_{\bf p'}} \Delta_{\bf p'} \,,
\eea 
which can be elegantly
derived via the Bogoliubov transformation\cite{Bogoliubov}. 
Here, $f(\epsilon)=(\exp(\epsilon-\mu)/T)+1)^{-1}$ is the Fermi distribution
function,  $E_{\bf k}=\sqrt{(\varepsilon_{\bf k}-\mu)^2+\Delta_0^2}$ the quasi-particle energy
and $\varepsilon_{\bf k}=\hbar^2k^2/2m$ the free particle energy.
In the BCS limit number conservation insures that $\mu=E_F$.  
$U({\bf p,p'})=U({\bf p-p'})$ is the Fourier transform of $U({\bf r})$.
In metals phonons provide a small residual attractive interaction $U({\bf p',p})\simeq -V_{ph}$
with a cutoff of order
the Debye frequency $\omega_D$. As result we obtain from Eq. (\ref{BCS})
the BCS s-wave gap $\Delta_0=2\hbar\omega_D e^{-1/N(0)V_{ph}}$ at zero temperature.

In the years immediately after BCS was developed the gap equation was generalized 
in terms of scattering lengths\cite{vanHove}. When the interaction $U({\bf r})$ 
is short range,
its Fourier transform $U({\bf p-p'})$ is long range in momentum. It is then convenient to
replace the interaction by its scattering matrix $T=U+UG_0T$, where
$G_0=1/(2\varepsilon_{\bf k}-i\delta)$, is the vacuum propagator for two particles.
At low momenta the scattering matrix is given by the s-wave scattering length
$T=4\pi\hbar^2a/m\equiv U_0$. 
Eliminating $U$ in the gap equation (\ref{BCS}) gives
\bea \label{Leggett}
  1=  \frac{U_0}{2V}\sum_{\bf k} 
    \left[\frac{1}{\varepsilon_{\bf k}}-\frac{1-2f(E_{\bf k})}{E_{\bf k}} \right] \,.
\eeq
Note that the difference between the vacuum and in medium Green's functions
automatically cuts off the high momenta, which are now 
included in the scattering length.
Solving this gap equation at zero temperature for a
Fermi gas interacting through an attractive scattering length $a<0$
gave a pairing gap in the dilute limit, $|a|k_F\ll 1$, \cite{vanHove}
\bea \label{dilute}
   \Delta_0 = \frac{8}{e^2} E_F
   \exp\left[\frac{\pi}{2ak_F}\right] \, . 
\eea

\subsection{Induced Interactions}

Gorkov\footnote{Inquiring into details about their calculation some 40
years later Gorkov only remembered
``...that it was a particular difficult calculation!''}  
pointed out that many-body effects 
(induced interactions) lead to the next order correction in the interaction
\cite{Gorkov,II,Martikainen,Kim}
\bea \label{ind}
  U_{ind}({\bf p'},{\bf p}) &=& -\frac{U_{12}^2}{M}
  \sum_{\bf q} \frac{f(\xi_1({\bf k'+q}))-f(\xi_1({\bf q}))} 
  {\xi_1({\bf k'+q})-\xi_1({\bf q})}  \nonumber\\
  &+& 
  \sum_{j,{\bf q}}\frac{U_{1j}U_{2j}}{M}
 \frac{f(\xi_3({\bf k+q}))-f(\xi_3({\bf q}))}
  {\xi_3({\bf k+q})-\xi_3({\bf q})}  .
\eea
$f$ is the Fermi distribution of $\xi_j({\bf q})=\epsilon_{\bf q}-\mu_j$,
${\bf k'}={\bf p}+{\bf p'}$ and ${\bf k}={\bf p}-{\bf p'}$,
where ${\bf p}_i$ are the momenta of the two pairing spin states. 
Induced interactions due to particle-hole loop diagrams 
from $j=3,...,\nu$ spin states are
responsible for the second sum in Eq. (\ref{ind}) and has opposite sign.
For two components, however, it is absent.

In the two-component spin-balanced system the induced interactions
effectively leads to
a second order correction to the scattering length 
$a\to a+(2/3\pi)\ln(4e)k_Fa^2$ \cite{Gorkov,II}, 
where $a=a_{12}$ and $U_{12}=U_0=4\pi\hbar^2a/m$.
Including this correction in Eq. (\ref{dilute}) 
reduces the gap by a factor $(4e)^{1/3}\simeq 2.2$
\bea \label{Gorkov}
   \Delta = \left(\frac{2}{e}\right)^{7/3} E_F
   \exp\left[\frac{\pi}{2ak_F}\right] \, . 
\eea

The induced interactions consists of a repulsive direct part and an
attractive part due a loop diagram which is therefore proportional to 
the number of components $\nu$.
The induced interactions therefore scale as $(3-\nu)$ in a system
of $\nu$ spin balanced multi-components with the same scattering length.
The corresponding gap is
\bea
  \Delta = (4e)^{\nu/3-1}\Delta_0 .
\eea
For spin polarized systems or for the $^6$Li systems with three
different scattering lengths the pairing gap is described in Refs. \cite{Martikainen}.
Adding bosons enhance pairing \cite{II}.

In the unitary limit $k_F|a|\ga 1$ the gap is of order the Fermi
energy \cite{Leggett}.  Extrapolating (\ref{Gorkov}) to
$ak_F\to\pm\infty$ \cite{long} gives a number $\Delta=0.49E_F$ close
to that found from odd-even staggering binding energies
$\Delta=0.54E_F$ calculated by Monte Carlo \cite{Carlson}.  The
crossover model of Leggett described below gives a somewhat larger gap
$\Delta^{Leggett}/E_F=0.69$.  Such values of order the Fermi energy
are one or two orders of magnitude larger those found in metals and
high temperature superconductivity and was met by disbelief among
condensed matter physicist and the crossover was considered academic.
Only after the realization and confirmation of the BCS-BEC crossover
in experiments with ultracold atoms was its significance acknowledged.

It also follows from the BCS gap equation that the gap and thus superfluidity and 
superconductivity all vanish at a critical temperature
\bea 
   T_c=\frac{e^{C_E}}{\pi}\Delta \simeq 0.567\Delta  , 
\eea
for any weak interaction ($C_E=0.577...$ is Euler's constant).
In the unitarity and BEC limits the critical temperature is no longer proportional
to the gap.

\begin{figure}[t]
\includegraphics[scale=.65,angle=-90]{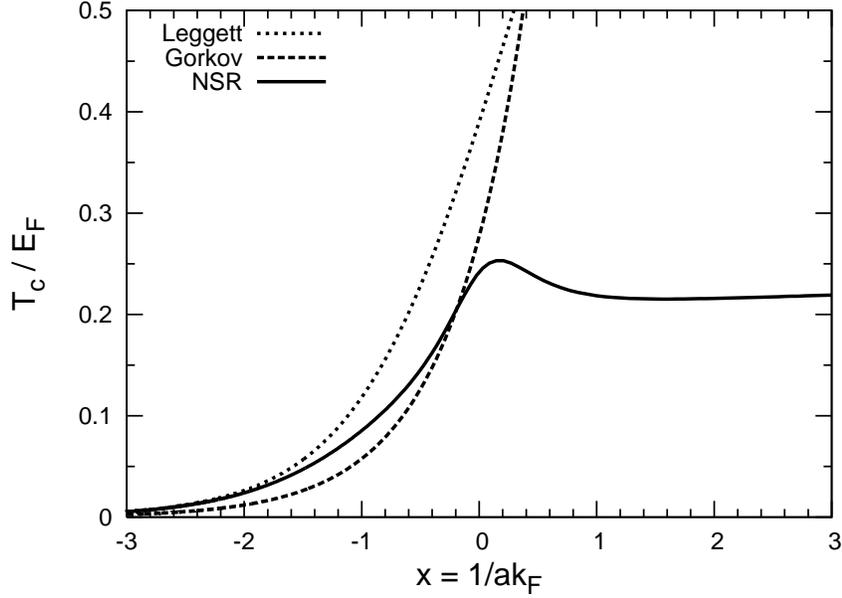}
\caption{Superfluid transition temperatures from the Leggett crossover, Gorkov and NSR.}
\label{FigTc}
\end{figure}

\subsection{Leggett's BCS-BEC Crossover}

Eagles and Leggett made the important step of connecting BCS pairing to a BEC of
molecular bosons via crossover.
Eagles gave an early treatment of the crossover from BCS superconductivity to a BEC in
the context of systems with low carrier concentrations \cite{Eagles}.
For pedagogical reasons we will follow Leggett who solved the 
gap equation of Eq. (\ref{Leggett}) and number conservation
\beq \label{numberconservation}
   N =\sum_{\bf k}\left[1-\frac{\varepsilon_{\bf k}-\mu}{E_{\bf k}} \right] \,, 
\eeq
self consistently in the crossover. 
The beauty of Leggett's crossover model Eqs. (\ref{Leggett}) and (\ref{numberconservation})
is that it describes the crossover continuously as function of the parameter
$x=1/ak_F$ from the BCS limit ($x\to-\infty$) over the unitarity limit $(x=0)$ to
the BEC limit $(x\to+\infty)$.
It is important to include (pairing) interactions in the chemical potential through
number conservation as it changes the chemical potential dramatically.

In the dilute (BCS) limit the gap equation leads to the standard BCS gap 
of Eq. (\ref{dilute}) - not including the Gorkov correction.
The chemical potential is
$\mu=E_F$ and does not include the standard
mean field Hartree-Fock correction of a dilute gas.

In the BEC limit the pairing gap approaches 
\bea
  \Delta=4E_F/\sqrt{3\pi ak_F} .
\eea
The chemical
potential approaches half of the molecular binding energy $\epsilon_M=-\hbar^2/ma^2$,
\bea \label{muLeggett}
 \mu = -\frac{\hbar^2}{2ma^2} + \frac{\pi\hbar^2na}{m} ,
\eea
plus the BEC mean field corresponding to a molecular scattering length
of $a_M=2a$. Four-body \cite{Petrov}, Monte Carlo calculations 
\cite{Casulleras} and experiments \cite{Ketterle} 
do, however, indicate that $a_M\simeq 0.6a$

On the BCS side the minimum quasiparticle energy is
$\Delta$ and occur when $k=k_F$. On the BEC side 
the chemical potential
is negative and the minimum quasiparticle excitation energy is the
quasiparticle energy for ${\bf k}=0$
\beq \label{Leg}
  E_{qp} = \sqrt{\mu^2+\Delta^2} \,.
\eeq
The quasiparticle energy is observed in the
spin excitation response function \cite{Burovski,Bloch}.

In the Leggett model the gap vanishes in the BEC limit at a critical temperature
$T_c^{dissoc}=|\epsilon_M|/\ln(|\epsilon_M|/E_F)^{3/2}$, which is smaller than the
quasi-particle excitation energy. It is not the on-set temperature for
superfluidity but rather a molecular pair dissociation temperature
\cite{NSR}. 
The onset of superfluidity occurs at a lower critical
temperature for a molecular BEC,
$T_c^{BEC}$.

The Leggett model fails in the BEC limit because of the basic
assumption that only opposite momenta fermions (zero total momentum pairs)
can pair. Pairs with non-zero momenta are thermal excitations
of molecular bosons and including such degrees of freedom lowers the 
critical temperatures increasingly towards the (molecular) BEC limit.

\subsection{NSR}

 The model of Nozi\'eres \& Schmit-Rink (NSR) \cite{NSR} extends the
Leggett model so that it correctly describes
the critical temperature in the molecular BEC limit.
We will therefore give a brief outline of NSR
with emphasis on how pair motion is included and how it corrects the Leggett model
in the BEC limit. Also, the NSR approach is formulated such that it applies to
optical lattices discussed in sec. (\ref{OL}) with few but crucial differences.

 In the NSR model the (molecular) pair momentum
${\bf q}$ is included in the
two-particle correlation function. To lowest order it is given by
the propagator for two free atoms
\bea
\Pi({\bf q},\omega_\nu) &=& \frac{T}{V} \sum_{{\bf k},\omega_m}
     G_0({\bf k},i\omega_m)G_0({\bf q-k},i\omega_\nu-i\omega_m) \nonumber \\  \label{Pi}
    &=& \frac{1}{V}\sum_{{\bf k}} 
       \frac{1-f(\epsilon_{{\bf q}/2+{\bf k}})-f(\epsilon_{{\bf q}/2-{\bf k}})}
         {\omega_\nu+2\mu-\epsilon_{{\bf q}/2+{\bf k}}-\epsilon_{{\bf q}/2-{\bf k}}} ,
\eea
 In Eq. (\ref{Pi}) 
the Matsubara frequencies $\omega_m=m2\pi Ti$ have been summed over integers $m$.

We now scale the potential by $\lambda$ and sum
up interactions energies $(U\Pi)^n$ from $n=1,2,...$ ladders.
We obtain for the expectation value of the interaction energy in state $\lambda$
gives 
\bea
   \langle \lambda U\rangle=-T \sum_{{\bf q},\omega_\nu} 
        \frac{\lambda U\Pi({\bf q},\omega_\nu)}{1-\lambda U\Pi({\bf q},\omega_\nu)} .
\eea
From the Hellmann-Feynmann theorem we now obtain the interaction part of the free energy
\bea \label{Free}
 \Omega_{int} = \int_0^1 \frac{dx}{\lambda} \av{\lambda U}
  = T \sum_{{\bf q},\omega_\nu} \ln[1-U\Pi({\bf q},\omega_\nu) ] .
\eea
As in the gap equation (\ref{Leggett}) large momenta contributions are removed by
replacing $U\Pi$ by $U_0\Pi_r$, where the renormalized propagator is
\bea
 \Pi_r({\bf q},\omega_\nu)=\Pi({\bf q},\omega_\nu)+\frac{1}{V}
 \sum_{\bf k}\frac{1}{2\epsilon_{\bf k}} . 
\eea
Inserting into Eq. (\ref{Free}) gives the
thermodynamic potential
\bea \label{Omega}
 \Omega = \Omega_0 + T \sum_{{\bf q},\omega_\nu} \ln[1-U_0\Pi_r({\bf q},\omega_\nu) ] ,
\eea
where $\Omega_0=-2T\sum_{\bf q}\ln[1+e^{(\epsilon_q-\mu)/T}]$ is the free
energy for non-interacting Fermions.
The frequency sum in Eq. (\ref{Omega}) can, using the residue theorem, be converted
to an $\omega$-integral of the Bose distribution function 
$(\exp(\omega/T)-1)^{-1}$ around
the real axis, where the logarithm in Eq. (\ref{Omega}) has a cut.

In the BEC limit, where $\mu$ is large and negative, it follows from Eq. (\ref{Pi})
that $1-U_0\Pi_r({\bf q},\omega)$ is proportional to 
$\omega -\epsilon_M(q)+2\mu$, where
\bea \label{EM}
   \epsilon_M(q) = -\frac{\hbar^2}{ma^2} + \frac{\hbar^2q^2}{4m} ,
\eea
is the molecular binding and kinetic energy of a pair with mass $2m$ and momentum ${\bf q}$.
Therefore the derivative of $\ln[1-U_0\Pi_r({\bf q},\omega_\nu)]$ with respect to $\mu$
has a pole at $\omega=\epsilon_M(q)-2\mu$ with residue 2.
Using the residue theorem again
we obtain the number equation
\bea \label{gapT}
   N \simeq -\frac{d\Omega}{d\mu} \simeq 2\sum_{\bf q} 
           \frac{1}{e^{(\epsilon_M(q)-2\mu)/T}-1} .
\eea
Here we have ignored the contribution to the thermodynamic potential $\Omega_0$ from free
Fermions
which is negligible in the BEC limit. The
number equation is simply that of a free Bose gas as opposed to that
of a Fermi gas Eq. (\ref{numberconservation}) in the BCS limit.

 Whereas the number equation changes qualitatively from free Fermi atoms to
free Bose molecules in the crossover, 
the gap equation is unchanged and given by Eq. (\ref {Leggett}).
In the BEC limit the gap equation simply yields that the chemical potential
is half the molecular binding energy, $\mu=-\hbar^2/2ma^2$. The number equation
now gives
\bea
  T_c=\frac{\pi}{[2\zeta(3/2)]^{2/3}}\frac{\hbar^2n^{2/3}}{m}\simeq0.218E_F ,
\eea
i.e., the critical temperature is independent of the pairing interaction
and given by the $T_c^{BEC}$ for the molecular BEC.
At $T_c^{BEC}$ all bosons are thermally excited with none remaining at zero momentum.

In general the gap equation (\ref{gapT}) and number conservation
$N=d\Omega/d\mu$ with $\Omega=\Omega_0+\Omega_{int}$
from Eq. (\ref{Free}) have to be solved self-consistently for
the critical temperature and the chemical potential 
in the BCS-BEC crossover. The result for $T_c$ 
is shown in Fig. (\ref{FigTc}). One notices that NSR predicts a maximum near the unitarity limit.
Recent experiments find $T_c/E_F=0.19\pm0.2$ \cite{Nascimbene} whereas earlier
found $T_c/E_F=0.29\pm0.3$ \cite{Thomas}.
This is compatible with the gap $\Delta=0.44E_F$ found in \cite{Schirotzek}.

Although the NSR model is a qualitative improvement of the Leggett
model by correctly describing $T_c$ in the BEC limit, it still omits a
number of effects such as particle-hole contributions, selfenergies
and induced interactions from the medium.  Consequently, the Hartree
field and Gorkov corrections are not included, and in the BEC limit
the molecule-molecule scattering length $2a$ implied by Eq.
(\ref{muLeggett}) is a factor $\sim 3$ too large \cite{Petrov}.  More elaborate
models \cite{Strinati,Bloch,Fuchs} include some of these effects
and do not find a maximum for $T_c/E_F$ around the
unitarity limit.

\section{Atomic traps and Nuclei}
\label{sec:7}

Traps are necessary for confining the atoms and creates a density distribution that
is maximal at central density and
decrease towards the cloud size $R$ where it vanishes.
Thus all densities are present at once which can be difficult to separate
experimentally in order to extract the (n,T) phase diagram.

From a nuclear point of view harmonic oscillator traps are wonderful toy systems since
they provide confined systems, which are a first
approximation to the nuclear mean field potential. Additionally we can
tune the interactions and therefore the level splitting and
pairing. Also millions of Fermi atoms can be confined and not just the
$\la 250$ neutron and protons in nuclei limited by fission. Therefore
one can study the crossover from few to infinite number of particles
which is necessary in order to e.g. link pairing in nuclei to that in
neutron and nuclear matter. Similarities between the (prolate) nuclear
mean field and cigar shaped optical traps can also be exploited in
studies of shell structures, pairing and collective modes.

\begin{figure}[t]
\includegraphics[scale=.75,angle=0]{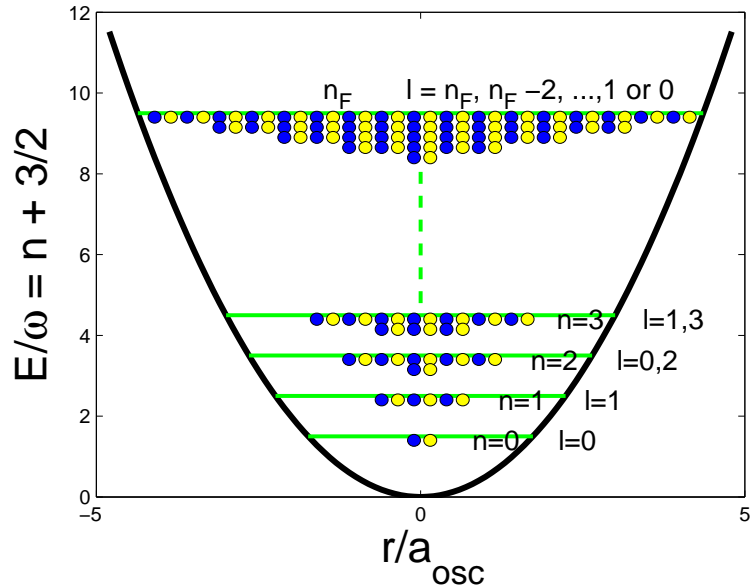}
\caption{Shells in a 3D harmonic oscillator trap.}
\label{FigHO}
\end{figure}

\subsection{Scaling in atomic traps}

In a noninteracting two spin balanced Fermi gas in a spherically symmetric HO potential
at zero temperature the particle fill HO shells of energy
$n\hbar\omega$, $n=1,2,..,n_F$ up to the Fermi shell $n_F=(3N)^{1/3}$ with
Fermi energy $E_F=n_F\hbar\omega$.
For a sufficiently large number $N$ of particles confined in a (shallow) trap
the system size $R$ is so long that
density variations and the extent of possible phase transition interfaces can be
ignored and
one can apply the local density and the Thomas-Fermi approximation.
Here the total chemical potential is given by the sum of the
harmonic oscillator (HO) trap potential and the local chemical potential 
$\mu_i=dE/dN_i$
\bea \label{TF}
  \mu_i(r) +\frac{1}{2}m\omega^2 r^2 
      = \frac{1}{2}m\omega^2 R_i^2 \,,
\eea
which must be constant over the lattice for all components $i=1,2,..$.
It can therefore be set to its value
at the its edge $R_i$, which gives the r.h.s. in Eq.
(\ref{TF}). We shall mainly discuss spin balanced two component systems where
the chemical potential and radii are equal. 
The equation of state determines $\mu(n)$ in terms of
the universal function of Eq. (\ref{TF}). 

In the dilute limit and in the unitarity limits
$\mu=\xi\hbar^2k_F^2/2m$, where $\xi=1$ and $\xi=1+\beta(0)$ respectively.
In both cases Eq. (\ref{TF}) gives 
$n(r)=k_F^3(r)/3\pi^2=n_0(1-r^2/R^2)^{3/2}$, 
where $n_0=\xi^{-3/4}(2n_F)^{3/2}/3\pi^2a_{osc}^3$ is the central density
and $R=\xi^{1/4}\sqrt{2n_F}a_{osc}$ the cloud size; $a_{osc}=\sqrt{\hbar/m\omega}$
is the oscillator length. The attraction contracts the gas to a higher central
density such that $R\propto k_F^{-1}(0) \propto \xi^{1/4}$.
The total energy of the trapped gas is $E/N=(3/8)\xi^{1/2}n_F\hbar\omega$. 

For a more general equation of state
$P\propto n^{\gamma+1}$ or $\mu\propto n^\gamma$ with polytropic index $\gamma$ 
the equilibrium density is
$n_{eq}=n_0(1-r^2/R^2)^{1/\gamma}$,
where $R^2=2(\gamma+1)P_0/\gamma n_0m\omega^2$. $P_0$ and $n_0$ are the
pressure and density in the centre of the trap. 
In both the non-interacting and the unitary limit $\gamma=2/3$ whereas
$\gamma=1$ in the BEC limit and for a bose gas.

As suggested in \cite{Ho} it is convenient to measure the density along the
axial axis integrated over transverse cross section $n(z)=\int n(r)dxdy$.
From the Gibbs-Duhem equation $(dP/d\mu)_T=n$ using
Eq. (\ref{TF}) we obtain the pressure in the centre along the axial axis
by integrating over transverse coordinates
\bea 
  P(z) = \frac{m\omega_\perp}{2\pi} n(z) .
\eea
At the same time the chemical potential $\mu(z)=m\omega_z(R_z^2-z^2)$ is known
by measuring the size $R(z)$ of the cloud along the $z$-axis. 
Thus $P(\mu)$ can be measured and
the equation of state extracted at any temperature.

In a Fermi liquid we obtain from Eqs. (\ref{PF}) and (\ref{betaFL})
that the pressure w.r.t. that in a free Fermi gas is
\bea
  P(x,T)/ P_{FG}= 1+\beta(x)+
    \frac{5\pi^2}{8}\frac{m^*}{m} \frac{T^2}{E_F^2} .
\eea
Detailed measurements of the intensive variables ($T,P,\mu$) gives
$\beta(0)=-0.49(2)$ and $m^*/m=1.13(3)$ \cite{Nascimbene}.

\subsection{Collective modes}

Tickling the trapped atoms sets them into oscillations at certain
eigen-frequencies called collective modes. Such giant dipole and quadrupole
modes have been important for studying nuclei \cite{BM}.
The collective modes can be calculated from the equation of state
and the Euler equation
\bea
   mn\frac{\partial{\bf v}}{\partial t} =
   -\nabla (P +nm\sum_i \omega_i^2r_i^2) \,.
\eea
where ${\bf v}=\partial{\bf r}/\partial t$ is the local velocity. 
The last term is the gradient of
a generally deformed HO potential.
The Euler equation can be solved analytically for polytropic equation of states.
For spherical symmetric 3D traps one finds collective modes
at eigen frequencies $\omega_{\eta l}$ 
for a mode with $\eta=0,1,2,...$ nodes and angular momentum $l$ 
given by \cite{mode}
\bea \label{spherical}
  \frac{\omega^2_{\eta l}}{\omega_0^2} = l+2\eta[\gamma(\eta+l+1/2)+1] \,.
\eea  
Similar expressions exist for modes in deformed traps \cite{Stringari}
which have been measured
in detail \cite{Thomas,Bourdel}.

For attractive interactions the system is superfluid at
zero temperature and therefore 
irrotational with a quenched moment of inertia. As the 
temperature is increased above $T_c$ the moment of inertia increase to
rigid value and hereby the critical temperature $T_c=0.19(1)E_F$ is found
\cite{Riedl}.
Deformed system can be rotated but has a
quenched moment of inertia $I=\delta^2I_{rigid}$ \cite{Clancy}.

\begin{figure}[t]
\includegraphics[scale=.75,angle=-90]{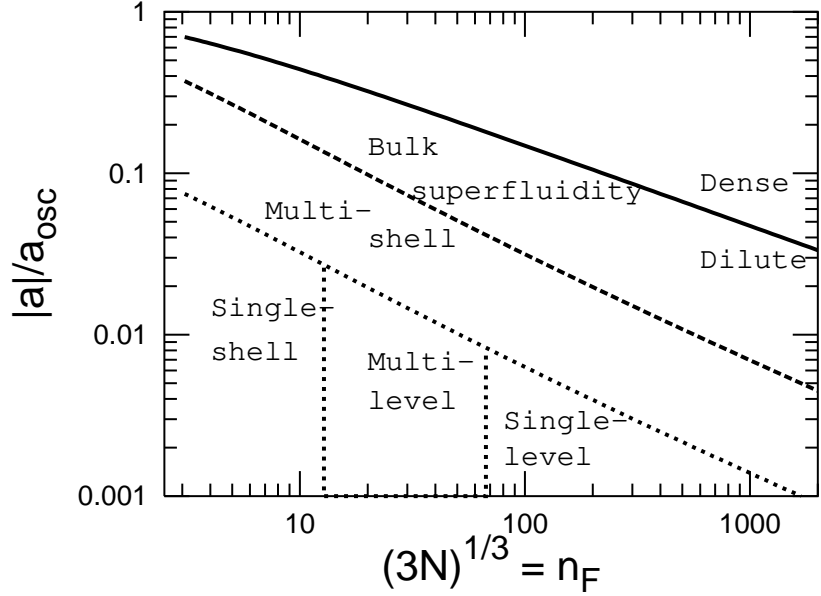}
\caption{Diagram displaying the regimes for the various pairing mechanisms
(see text) at zero temperature
in h.o. traps vs. the number of particles $N=n_F^3/3$
and the interaction strength $a$.
The dotted lines indicate the transitions between single-shell pairing 
$\Delta=G$, multi-level, single-level, and multi-shell pairing.
At the dashed line determined by $2G\ln(\gamma n_F)=\hbar\omega$ the pairing
gap is $\Delta\simeq\hbar\omega$, and it 
marks the transition from multi-shell pairing to bulk superfluidity
Eq. (\ref{Gorkov}). The pairing gap is $\Delta=0.54E_F$
above the full line $\rho|a|^3\ge 1$, which separates the dilute from 
the dense gas. From \cite{HM} }
\label{pairphases}
\end{figure}

\subsection{Shell structure and pairing in atomic traps}

The pairing gap generally increase with the level degeneracy or
density of states \cite{BM}.  For pairing in a single level of angular
momentum $l$ or shell $n_F$ the pairing gap scales with the number of
degenerate states $(2l+1)$ or $(n_F+1)(n_F+2)$ respectively. If levels are
split or distributed the gap depends on the density of levels around
the Fermi level as in the uniform case of Eq. (\ref{dilute}).

For very weak interactions the
level splitting of single particle levels $l=n_F,n_F-2,...,1$ or $0$ 
in the HO shell $n_F$ is smaller than the pairing gap when
there are sufficiently few ($n_F\la10$) atoms in a HO potential.
The pairing occurs between all states in the shell leading to
the supergap \cite{HM}
\bea
   G = \frac{32\sqrt{2n_F+3}}{15\pi^2} \frac{|a|}{a_{osc}}\hbar\omega \,.
\eea
For more particles or stronger interactions 
the situation becomes more complicated depending on the
gap size with respect to the level splitting as shown in Fig. (\ref{pairphases}).
Increasing the number of particles cause level splitting so that
pairing is reduced to multi-level and eventually single level pairing. 
Increasing the interaction
strength increase pairing to nearby shells \cite{HM} referred to as 
multi-shell pairing in Fig. (\ref{pairphases}). 
The general case can be solved within the Bogoliubov-deGennes 
equations \cite{HM,A}. For stronger interactions and many particles
the pairing approaches that in a uniform or bulk system, and eventually
dense system with $|a|k_F>1$ and the unitarity limit is reached.

The various pairing mechanisms and phases of Fig. (\ref{pairphases})
can be studied at low temperatures is which hopefully will be
reached in the near future.

\subsection{Pairing in nuclei}

We can exploit
some interesting similarities between pairing in nuclei \cite{BM,Pines}
and that of Fermi atoms in traps \cite{HM,A}.  To a first
approximation the nuclear mean field is often taken as harmonic
oscillator (HO) potential just as the optical traps.
Secondly, the residual pairing interaction between
nucleons is taken as a short range (delta function) interaction
as for cold atoms.

Before these result can be applied to nuclei there are, however, 
a number of differences that must be taken into account.
Large nuclei have approximately
constant central density $\rho_0\simeq 0.14$~fm$^{-3}$ and Fermi energy
$E_F$ in bulk. Therefore the
HO frequency, which is fitted to the nuclear mean field, 
decreases with the number of nucleons $A=N+Z$, where $N$ now
is the number of neutrons and $Z$ the number of protons in the nucleus, as
$\hbar\omega\simeq E_F/n_F \simeq 41 {\rm MeV}/ A^{1/3}$.
In the valley of $\beta$-stability the number of protons is
$Z\simeq A/(2+0.0155A^{2/3})$. Therefore in medium and large nuclei the difference
between the Fermi energies of
protons and neutrons exceeds the pairing gap so that pairing
between protons and neutrons does not occur.

Secondly, the nuclear mean field deviates from a HO potential by being 
almost constant inside the nucleus and vanish outside. The resulting net
anharmonic nuclear field is {\it stronger} and {\it opposite} in sign to the
corresponding (anharmonic) mean field in atomic traps. Therefore, the level
splitting is larger and the ordering
of the l-levels is reversed. In addition,
a strong spin-orbit force splits the single particle
states of total angular momentum $j=l\pm 1/2$, such that the
$j=n_F+1/2$ is lowered down to the shell ($n_F-1$) below.
The level splitting can be parametrized by a single parameter taken from analyses
of nuclear spectra. It increases with shell number up to $n_F\simeq 6$ for
heavy nuclei. 
Due to the
strong spin-orbit force the $j=l\pm 1/2$ states are split and the
$j=n_F+1/2$ is lowered down to the shell below.  The magic numbers become
$N,Z=8,14,28,50,82,126,184,..$, etc.  rather than the h.o. filled
shell particle numbers $N,Z=2,8,20,40,70,112,168,240,...$, etc.

\begin{figure}[t]
\includegraphics[scale=.75,angle=-90]{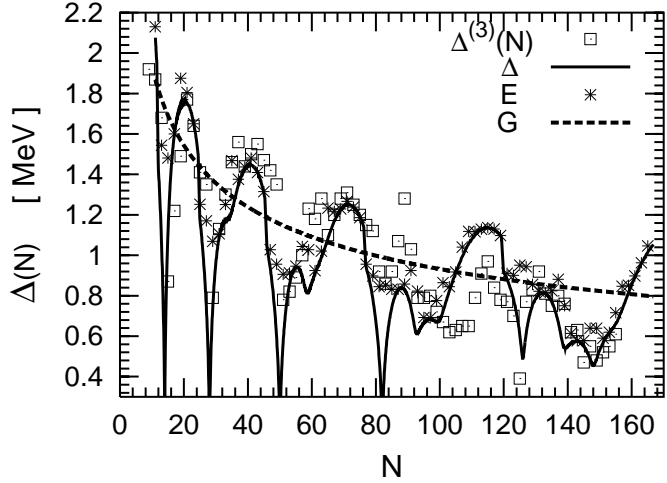}
\caption{Neutron pairing energies
vs. the number of neutrons. The experimental odd-even staggering energies
$\Delta^{(3)}(N)$ are averaged over
isotopes \protect\cite{Audi,Naza}. The calculated gaps $\Delta$ and
quasi-particle energies $E$  are obtained
from the gap equation (see text) with effective coupling strength 
$a=-0.41$~fm. The supergap $G$ is shown with dashed line. See
\cite{A} for details and corresponding plot for protons.}
\label{N}
\end{figure}

The pairing gaps and quasi-particle energies
can now be calculated by solving the Bogoliubov-deGennes gap equation \cite{HM}
which are shown in Fig. (\ref{N}) for neutrons.
The strong level splitting in nuclei has the effect that pairing is strongest when
the shell is half filled and weak near closed shells simply because there are fewer
states available for pairing, i.e., the level density is smaller.
Averaging over several shells, however, the mean gap is well approximated by
the supergap because the reduction in pairing due to level splitting is 
compensated by additional pairing to nearby shells.
Since $\hbar\omega$ scale as $\sim A^{-1/3}$
and $a_{osc}\propto n_F^{1/2}$ the single-shell
pairing gap also scales as $G\sim A^{-1/3}$. Therefore, the pairing gaps in light and
medium mass nuclei scale approximately as \cite{A}
\bea
  \Delta\simeq G\simeq \frac{|a|}{0.41{\rm fm}} \frac{5.5{\rm MeV}}{A^{1/3}} 
  \,.
\eea
As shown in Fig. (\ref{N}) the supergap
does not depend on the level-splitting and is therefore a 
robust prediction for the average magnitude and mass scaling of pairing gaps
in nuclei.

The data on neutron and proton pairing is obtained from the odd-even
staggering of nuclear binding energies $B(N,Z)$. It has been shown
that mean field contributions can be removed \cite{Naza} by using the
three-point filter $\Delta^{(3)}(N)= (-1)^N [B(N-1,Z)/2+B(N+1,Z)/2
  -B(N,Z)]$.  We compare in Fig. (\ref{N}) to the experimental
$\Delta^{(3)}(N)$ averaged over isotopes with the calculated gaps.
The analogous for protons $\Delta^{(3)}(Z)$ averaged over isotones
compare similarly to Bogolibov-deGennes calculations \cite{HM,A}. In
the calculations the effective coupling is the only adjustable
parameter which is fitted to the experimental data.  For both neutrons
and protons we extract $a\simeq -0.41$~fm.

Considering the simplicity of the model it describes a large number of
experimental gaps fairly well on average.  In a number of cases,
however, the calculated pairing gaps differ significantly from the
measured neutron gaps. Some of these deviations can be attributed to
the crude single particle level spectra assumed. 

The pairing in nuclear matter can now be estimated once the effective
interaction has been determined using Eq.  (\ref{dilute}). 
Note that the effective interaction includes induced interactions. Inserting
$a=-0.41$~fm and $k_F=1.3$~fm$^{-1}$ at nuclear saturation density,
$\rho_0=0.15$~fm$^{-3}$, we obtain the proton and neutron pairing gaps
\bea
  \Delta \simeq 1.1 {\rm MeV} \,, 
\eea
in the bulk of very large nuclei and in symmetric nuclear matter at
nuclear saturation density.  This number is compatible with earlier
calculations \cite{NS} of the $^1S_0$ pairing gap in nuclear and
neutron star matter around normal nuclear matter densities.

Neutron star matter has a wide range of densities and is
asymmetric, $Z/A\sim 0.1$, above normal nuclear matter densities. 
One can attempt to estimate of the pairing gaps as
function of density from the gap in bulk, Eq. (\ref{dilute}), with
$a\simeq -0.41$~fm and the neutron or proton Fermi wave numbers
$k_F^{N,Z}=(3\pi^2\rho_{N,Z})^{1/3}$ as function of densities. 
However, the effective interaction $a$ is
density dependent. At higher densities we expect that 
the effective interaction
becomes repulsive as is the case for the nuclear
mean field at a few times nuclear saturation
density. At lower densities the effective scattering length should approach
that in vacuum which for neutron-neutron scattering
is $a(^1S_0)\simeq -18$~fm.

In spite of several simplifying approximations in this atomic trap
model for nuclear pairing it provides at least a qualitative 
description of pairing in most nuclei. The effective interaction
 $a=-0.41fm$ is the same for both neutrons and protons which reflects that
both are close to forming bound states. At central densities
$x=1/ak_F\simeq -1.9$, i.e. on the BCS side of the unitarity limit.

\subsection{Quark and gluon matter}

The Coulomb and QCD interactions $\sim g^2/r$ or their Fourier transform
$\sim g^2/q^2$ are long range and therefore contrary to the short range
interactions underlying universality.
The long range QED and QCD interactions requires screening of infrared divergences \cite{EU}
whereas in unitary gases ultraviolet cutoffs are provided by renormalizing
the short range interaction in terms of the scattering length.
Never the less a number of similarities have appeared.

Confinement is caused by strong color fields between quarks and gluons
which forms bound state hadrons at temperatures below $T\la
T_{QGP}\simeq 160$MeV and densities below a few times normal nuclear matter
density, $\rho_0=0.15$fm$^{-1}$. These may again form a nuclear liquid at
temperatures below the nucleon gas critical temperature $T\la
T_{NG}\simeq 15$MeV for a four component nuclear system as discussed
above.  In the hadron gas the interparticle distance is larger than
the confinement interaction range, and the quark molecules can
therefore be viewed as a molecular gas. However, color neutrality
requires either Fermi molecules of three quarks (baryons) or
quark-antiquark Bose pairs (mesons). Of the two phase transitions in
the molecular gas of hadrons only the neutron gas can be viewed as a
BCS-BEC crossover as discussed in sec. (\ref{instability}).  The
instability and first order transition of a four-component nuclear gas
was discussed in sec. (\ref{instability}) can be viewed as
multi-component crossover.  Here the effective scattering length is
that between neutrons and protons $a\sim 20$fm, i.e. on the molecular
BEC side, and the nuclear matter is in the unitarity limit in the
sense that $x=1/ak_F\sim 0_+$ although they now are kept in place by
short range repulsive forces.  The transition from nuclear to quark
matter or a hadron gas to a quark-gluon plasma is quite different
because the interaction range is always of order the interparticle
distance. The strong short range repulsive forces between nucleons
leads to a strongly correlated wavefunction as in hard sphere
scattering and the strongly correlated nuclear fluid. 
In this case the Gutzwiller approximation of
Ref. \cite{Zhai} is valid and the matter does not undergo a
ferromagnetic transition separating the unlike components 
(see Fig. (\ref{FigPhases})). 
As seen from the
quark-gluon side one can define a ``crossover parameter'' $x\sim
\Lambda_{QCD}/k_F$ which at high densities is in the unitarity
limit. With decreasing densities the running coupling constant
$\Lambda_{QCD}$ diverges when the quark-gluon plasma undergoes a first
order transition to a hadron gas with $x\ga1$.

Quark pairing also have similarities to pairing in nuclear and
multi-component atomic gases. Color superconductivity between the 2
spin, 8 color and 2-3 flavor states is very sensitive to flavor
imbalance \cite{Alford}.  If the strange quark mass is small the up,
down and strange quark Fermi levels are close. If pairing is
sufficiently strong such that the gap exceeds the Fermi level
splitting all flavors can pair.  It is amazing that properties of
pairing in quark matter, which may exist in unaccessible cores of
neutron stars, can be studied in tabletop experiments with ultracold
multicomponent spin-imbalanced Fermi atomic gases.

Another interesting similarity is elliptic flow. The overlap zone in
semi-central high energy nuclear collisions is prolate (cigar shaped)
as are optical traps.  In subsequent expansion the hydrodynamics
forces stronger expansion in the direction where systems is narrowest
initially. This makes the momentum distribution azimuthally asymmetric
- referred to as elliptic flow. Results indicate that both ``liquids'',
the QGP/hadronic and ultracold unitary atomic gas,
expand as almost perfect fluids initially i.e. their viscosities are
record breaking low \cite{viscosity}.

Correlations can reveal the quantum phase structure. Originally
Hanbury-Brown \& Twiss measured the Bose-Einstein correlations between
stellar photons and determined the diameters of nearby stars.  Similar
correlations between mesons in high energy nuclear collisions have
been exploited to determine the freezeout size of the collision zones
\cite{HBT}.  Also Fermi anticorrelations have been observed between
baryons.  In ultracold atomic systems analogous ``noise'' correlations
have been found near the BCS-BEC crossover due to pairing
\cite{Ketterle}.  Bragg peaks have been observed for bosons in 3D
\cite{Folling} and 2D \cite{Spielman} lattices, and dips for 3D
fermions in \cite{Rom}.

\section{Optical Lattices}
\label{OL}

\begin{figure}[t]
\includegraphics[scale=.4,angle=0]{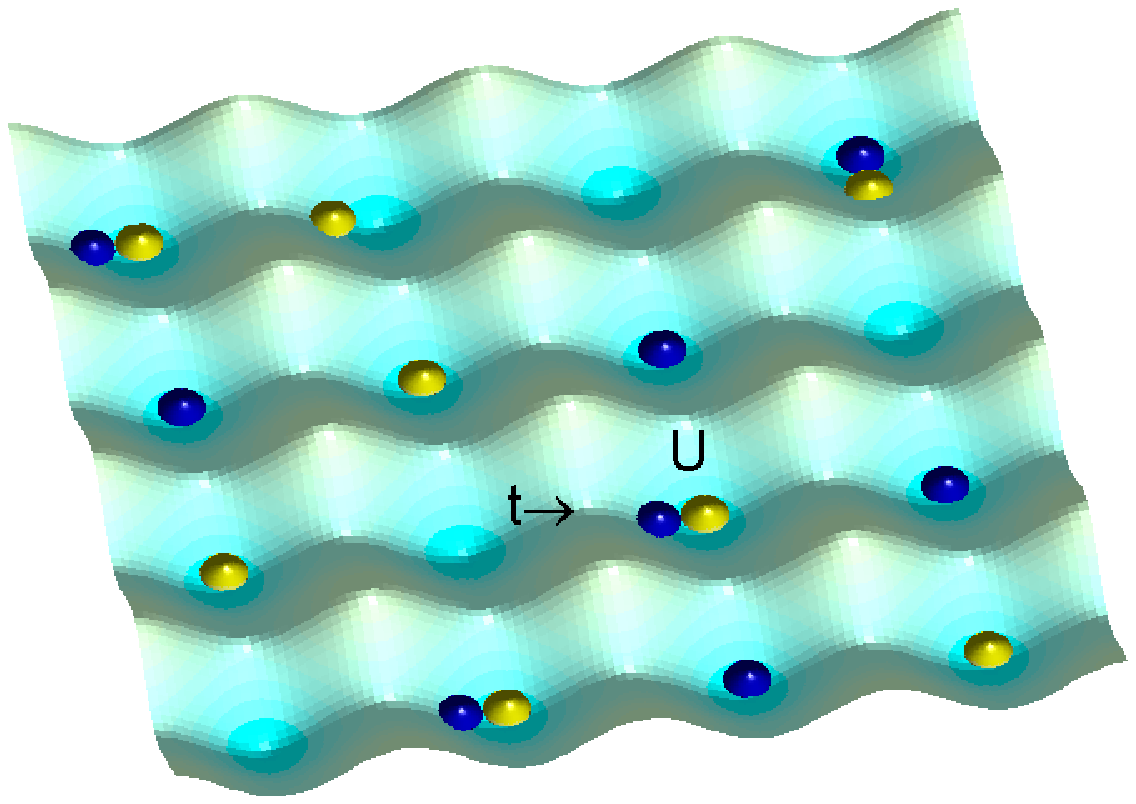}
\includegraphics[scale=.4,angle=0]{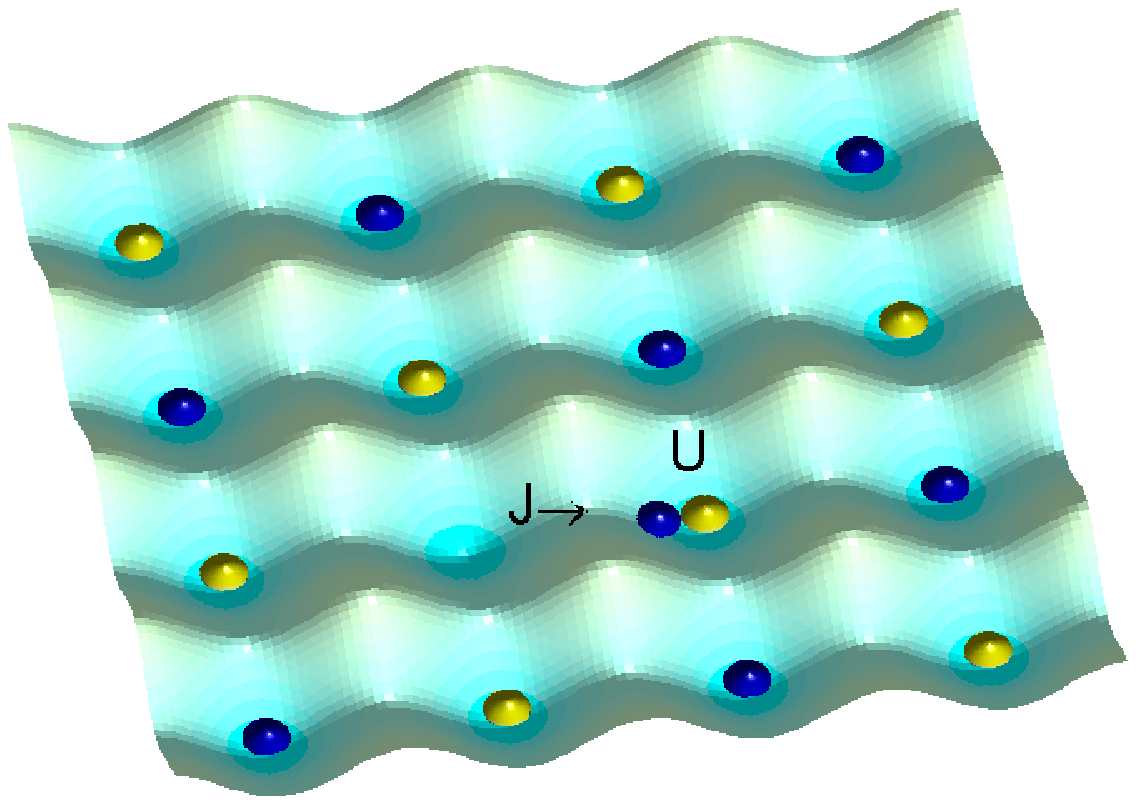}
\caption{2D optical lattices loaded with fermions in two
spin states (yellow and blue) at half filling. For onsite attraction
($U<0$, left) opposite spins tend to pair. For onsite repulsion
($U>0$, right) the ground state is an AF with virtual hopping
$J=4t^2/U$.}
\label{2DOL}
\end{figure}

A standing wave can be generated by shining lasers onto traps which
generates a 3D cubic or 2D square lattice potential with spacing
$\lambda=2\pi/k$, which is half the laser wavelength. The lattice
height $V_0$ varies with laser intensity and the scattering length.
In the tight binding approximation the onsite coupling
\bea
   U=E_Rak\sqrt{8/\pi}\xi^3
\eea
and hopping parameter
\bea
   t=E_R(2/\sqrt{\pi})\xi^3e^{-2\xi^2} ,
\eea 
where $E_R=\hbar^2k^2/2m$ is the recoil energy and $\xi=(V_0/E_R)^{1/4}$.
Varying $V_0$ and the scattering length $a$ near Feshbach resonances allows us
to tune the Hubbard parameters $U$ and $t$.

The lattice constant $\lambda$ introduces another lengths scale so that
the thermodynamic quantities generally depend on both density and interaction
strength. Therefore universality is broken except for low filling which corresponds
to dilute Fermi gases.

The Hubbard Hamiltonian on a D-dimensional lattice
describes optical lattices that are sufficiently deep for a one-band model
to apply,
\bea \label{Hubbard}
 H &=& \sum_{i,\sigma<\sigma'} U_{\sigma,\sigma'}
       \hat{n}_{i\sigma} \hat{n}_{i\sigma'}
-t\sum_{\av{ij},\sigma} \hat{a}_{i\sigma}^\dagger \hat{a}_{j\sigma} \,.
\eea
Here $\hat{a}_{i\sigma}^\dagger$ is the Fermi 
creation operator of the hyperspin states $\sigma=1,2,...,\nu$, 
$n_{i\sigma}=\hat{a}_{i\sigma}^\dagger \hat{a}_{i\sigma}$ the density
and $\av{ij}$ denotes nearest neighbours with hopping parameter $t$.
Due to particle-hole symmetry results for a given site
filling $n$ also applies to $2-n$.

The Hubbard model is of fundamental importance in condensed matter physics
where it explains properties of Mott insulators, antiferromagnets,
d-wave superconductivity, etc. The Hubbard model can only be solved in 1D for
fermions but a number of models and numerical calculations have given insight
in the various phases and order parameters. Optical lattices may tell us
the answers in a few years.

\subsection{On-site attraction and Pairing}

We start by investigating s-wave pairing between to spin balanced states due to an
attractive on-site interaction $U<0$. 
The mean field gap equation for singlet superfluidity 
at zero temperature is \cite{Micnas} 
\bea \label{sgap}
 \Delta_{\bf p'} = -\frac{1}{M} \sum_{\bf p} U({\bf p'},{\bf p})
    \Delta_{\bf p} \frac{\tanh(E_{\bf p}/2T)}{2E_{\bf p}} \,,
\eea
where $M$ is the number of lattice points and
$E_{\bf p}=\sqrt{(\epsilon_{\bf p}-\mu)^2+\Delta_{\bf p}^2}$ with
$\epsilon_{\bf p}=2t\sum_{i=1,D}(1-\cos p_i)$.

The density 
$n=1-\sum_{\bf p}(\epsilon_{\bf p}/E_{\bf p})\tanh(E_{\bf p}/2T)/M$, is
also the filling fraction in units where the lattice constant is unity.
Momenta are in the first Brillouin zone only $|p_i|\le\pi$ taking
$\lambda=1$ for convenience.

Due to particle-hole symmetry results also apply replacing the density
by $(2-n)$ and chemical potentials by $(4Dt-\mu)$, where $4Dt$ is
the bandwidth in $D$ dimensions. At low filling $\epsilon_p=tp^2$ and
results also apply to uniform systems replacing $t=1/2m$.

\begin{figure}[t]
\includegraphics[scale=.65,angle=-90]{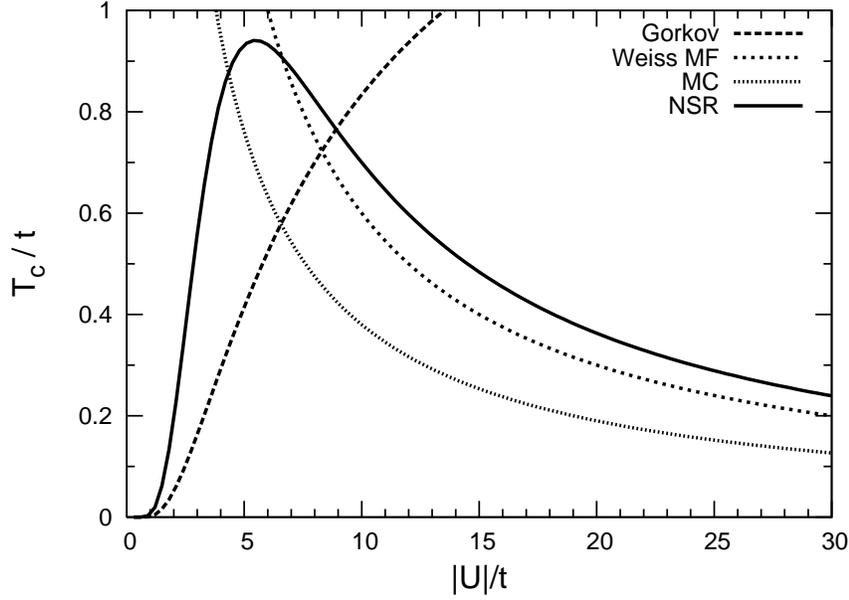}
\caption{Superfluid transition temperatures in a 3D lattice
at half filling for the NSR model,
the Heisenberg model within (Weiss) mean field $T_c=1.5J$ and Monte Carlo.
``Gorkov'' is the weakly attractive limit of Eq. (\ref{G3D1}) reduced by
a factor 3.9 due to induced interactions.}
\label{FigTcl}
\end{figure}

\subsubsection{Pairing in the dilute limit}
We first calculate the zero temperature gaps for weak attraction
$|U|\ll t$ excluding induced interactions. The s-wave gap
$\Delta_{\bf p'}=\Delta_0$ is then momentum-independent because the
interaction $U$ is.  The gap equation (\ref{sgap}) reduces to: 
$1=-(U_0/M)\sum_{\bf p}1/2E_{\bf p}$, from
which the gap can be calculated analytically at low and half filling 
to leading orders in 3D and 2D (for 1D see, e.g. \cite{Marsiglio}).

The 3D lattice has a critical coupling \cite{Burovski}
\bea
   U_c=-M/(\sum_{\bf q} 1/2\epsilon_{\bf q})
=-8\sqrt{2}t/[\sum_{l=0}^\infty P_{2l}(\sqrt{9/8})(2l-1)!!/2^{2l}l!]
\simeq -7.913t ,
\eea
where a two-body bound state can be formed at zero
density ($\mu=0$). This threshold corresponds to the unitarity limit of
infinite scattering length in the uniform system
and naturally enters the sum in the gap equation.
The dilute gap becomes to leading orders
\bea \label{G3D}
  \Delta_0^{3D} = \frac{8}{e^2} \mu \exp\left[\frac{4\pi^2 t}{k_F}
   \left(\frac{1}{U}-\frac{1}{U_c}\right) \right] ,\, n\ll1 ,
\eea
at low densities where $\mu=tk_{F}^2\ll t$.
The level density at the
Fermi surface $N(\mu)=k_{F}/(4\pi^2t)$ enters in the exponent as in standard
BCS theory. $U_c$ acts as a cutoff in
the one-band Hubbard model which in the uniform (continuum) limit 
is absorbed into the scattering length $a$.
Thus Eq. (\ref{G3D}) is the finite lattice equivalent of the
gap in the uniform system $\Delta_0^{3D}=(8/e^2)\mu\exp(\pi/2k_{F}a)$,
with $U^{-1}-U_c^{-1}=U_0^{-1}$, where $U_0=4\pi a/m$ as in the uniform
case, such that
threshold $U=U_c$ corresponds to $|a|=\infty$. 

Near half filling the level density is surprisingly constant
$N(\mu)\simeq 0.143/t$ in a wide range $4t\le\mu\le8t$ around half filling. 
To leading orders the gap becomes \cite{Micnas}
\bea\label{G3D1}
  \Delta_0^{3D} = \alpha t \exp\left[\frac{1}{N(\mu)U}\right] ,\, n\simeq 1,
\eea
where the prefactor $\alpha\simeq6.544$ can be calculated numerically.

The 2D lattice has a superfluid s-wave gap that be calculated
from the gap equation \cite{Dupuis}
\bea \label{G2D}
  \Delta_0^{2D} = \sqrt{8\pi\mu t} \,\exp\left[\frac{4\pi t}{U}\right] \,,
  \, n\ll 1,
\eea
when the density $n=k_F^2/2\pi$ is small. 
The 2D level density is $N(\mu)=1/(4\pi t)$.
Eq. (\ref{G2D}) assumes that the
on-site coupling and the gap are small such that 
$\Delta_0^{2D}\la\mu$. This is not fulfilled at sufficiently
low densities where instead $\Delta_0^{2D}=4\pi t\exp(8\pi t/U)$, 
which also is the two-body binding energy in 2D at zero density.
There is always a two-body bound state in 2D with purely attractive
interaction and therefore $U_c$ vanishes. In 3D a similar pair condensate
(a molecular BEC) 
appears when $a>0$ corresponding to $U<U_c$.

For intermediate fillings the pairing is
more complicated in 2D. Near half filling $n\simeq1$ the level density
$N(\epsilon)=\ln(16t/|\epsilon-4t|)/(2\pi^2t)$
has a logarithmic singularity due to the van Hove singularity.
Calculating the r.h.s. of the gap equation therefore gives a double log:
$1=|U|/(4\pi^2)\ln^2(32t/\Delta_0^{2D})$, to leading logarithmic orders,
resulting in the gap when $n\simeq 1$ \cite{Dupuis}
\bea
 \Delta_0^{2D}= 32t \exp(-2\pi\sqrt{t/|U|}) \,.
\eea

In the strong coupling limit $\Delta=|U|/2$.

\subsubsection{Induced interactions on the lattice}

In 3D the induced interactions are at low densities $n\ll1$ the same as in the
uniform case Eq. (\ref{ind}) and reduce gaps and $T_c$ by a factor
$(4e)^{1/3}\simeq 2.2$. Near half filling the reduction factor is $\sim 3.9$
for weak interactions. For stronger interactions the
induced interactions are enhanced \cite{Martikainen} leading to a
reduction of the gap by a factor $\sim25$ when $U=-3t$. 

 In 2D the
induced interactions suppress the gap by a factor $e$ in the dilute
limit \cite{Nishida} but are divergent at half filling due to the singular
level density \cite{Martikainen}.
The induced interactions have been calculated for general spin polarization
and number of components in 3D and 2D optical lattices \cite{Kim}.

\subsubsection{Crossover and the BEC limit}

 As the interactions become stronger the Fermi gas gradually undergo
crossover to a molecular BEC. On the lattice, however, the finite
bandwidth automatically provides a cutoff which affects the crossover and
changes the BEC limit on the lattice qualitatively from the uniform case. 
We shall describe this crossover
employing the NSR model as above for the uniform case but with
a finite bandwidth. 
Our starting point is again the two-particle correlation
function of Eq. (\ref{Pi}). In the BEC limit (large negative chemical
potential) we find from Eq. (\ref{Pi}) that $1-U\Pi({\bf q},\omega)$ is proportional to
$\omega-\epsilon_M(q)+2\mu$ as in NSR, where now the molecular boson energy is
\bea \label{EL}
  \epsilon_M({\bf q}) = -U + J\sum_{i=1,3} [1-\cos(q_i)]
\eea
with $J=4t^2/U$. 
Therefore the number equation reduces to the condition for BEC in an ideal
Bose gas as Eq. (\ref{gapT}) 
\bea \label{gapN}
   N \simeq -\frac{d\Omega_{int}}{d\mu} \simeq 2\sum_{\bf q} 
   \frac{1}{e^{(\epsilon_M({\bf q})-2\mu)/T}-1} ,
\eea
but with the molecular boson energy of Eq. (\ref{EL}) in stead of (\ref{EM}).
The gap equation at $T_c$ yields a chemical potential $\mu=-U/2$ in the BEC limit,
and inserting this in the number equation gives a critical temperature
\bea
  T_c=\frac{2\pi}{[2\zeta(3/2)]^{2/3}}n^{2/3} J \simeq2.09n^{2/3} J ,
\eea
when $n\ll1$. For $n=1$ we obtain $T_c=1.71 J$.

The crossover in optical lattices is shown in Fig. (\ref{FigTcl}) at half filling.
The critical temperature decreases with interaction strength
in the BEC limit and thus
differs from uniform systems where it approaches a constant $T_c^{BEC}$.

The Hubbard model can in the BEC limit be mapped onto the
Heisenberg spin model in a magnetic field for which accurate calculations exist, which
allows us to check the NSR and other models \cite{Tamaki}.
Half filling corresponds precisely to zero magnetic field, i.e.,
\bea
   H = J\sum_{\av{ij}} {\bf S}_i\cdot {\bf S}_j ,
\eea
and the N\'eel temperature corresponds to $T_N=T_c$ in the BEC limit.
In (Weiss molecular) mean field theory $T_N=1.5J$, quantum Monte Carlo
$T_N=0.95J$, high temperature expansions $T_N=0.90J$, as compared to
$T_c=1.71J$ in NSR. Extending the NSR model with charge and spin density fluctuations
gives a smaller value $T_c\simeq0.4J$ because molecular repulsion is overestimated
\cite{Tamaki}.

\subsection{On-site repulsion}

On-site repulsive interactions $U>0$ generally disfavours doubly occupied sites
and lead to a Mott gap at half filling of order $\sim U/2$ at very low temperatures. 
Mott insulator (MI) transitions are observed for bosons in traps at fillings $n=1,2,3,4,5,..$
\cite{Ketterle}. For fermions the phase diagram is more complex, an antiferromagnet (AF)
at and near half filling, a paramagnet (PM) otherwise and possible a ferromagnet
for very strong repulsion \cite{Pruschke}. Furthermore the AF phase may be unstable towards phase
separation to two coexisting phases: an AF at half filling and a PM at filling
slightly below or above half filling. The co-existing phase may also be unstable towards
stripes. In 2D these phase also compete with d-wave superfluidity.

\subsubsection{Antiferromagnetism}
At half filling fermions are known to form a MI in 1D \cite{Lieb} whereas
in two and higher dimensions an antiferromagnetic (AF) insulator is found
in Monte-Carlo calculations. The AF alternating spin order on the lattice is driven by the
simple fact that hopping can occur to a neighbouring site only if it is occupied by
an opposite spin, which generates a
super-exchange coupling.

We will here study the AF phase and its transition to 
a paramagnet at the N\'eel temperature $T_N$. Remarkably, it undergoes a crossover
very similar to the critical temperature described above for attractive interactions,
i.e. $T_N\sim T_c$.
For weak repulsion $T_N$ can be determined from
the mean field gap equation
\bea
   1 = \frac{U}{M} \sum_{\bf p} \frac{\tanh(\epsilon_{\bf p}/2T_N)}{2\epsilon_{\bf p}} ,
\eea
which is identical to that for $T_c$ of Eq. (\ref{sgap}) except for the sign of $U$.
Therefore $T_N=T_c$ within mean field. Induced interactions can be included
as for $T_c$ as described above and therefore reduce $T_N$ by a factor  $\sim3.7$ at
half filling \cite{Dongen} which is very similar to the reduction of $T_c$ by
induced interactions as discussed above.

For stronger coupling fluctuations reduce the mean field gap and $T_N$
as for the NSR model of $T_c$. At very strong coupling the repulsive
Hubbard model can at half filling also be mapped onto the Heisenberg
model with coupling $J$. Therefore $T_N$ is also in this limit given
by $T_c$ as function of $|U|$, and the details of the crossover
\cite{Staudt,Georges} are quite similar to the BCS-BEC crossover.

The AF phase transition masks a possible MI phase. 
By frustrating the system one can, however, remove the AF order and
observe the MI transition by a vanishing conductivity or compressibility.
Another way to suppress the AF order is to include multi-components which naturally
upsets the alternating spin up/down order in a two-component AF. The limit of
many components actually correspond to a Bose system where MI transitions are found
at every integer filling.

\subsubsection{Phase separation}

Mean field theory provides a first impression of the phases competing for the
ground state and has the advantage that it is computationally simple
as compared to more complicated theories.  The MF equations for the
Hubbard model are standard and we refer to, e.g., Refs. \cite{Micnas,Langmann}. The
energy densities can be calculated within the Hartree-Fock
approximation for the paramagnetic (PM), ferromagnetic (FM), antiferromagnetic
(AF) and phase separation (PS). 

At low density $n\ll 1$ the ground state is that of a dilute paramagnetic (PM) gas
with energy
\bea \label{PM}
  \varepsilon_{PM} = -4tn+\left[\frac{\pi}{2}t+\frac{1}{4}U\right]n^2 
 + {\cal O}(n^3)\,.
\eea
In units where the lattice spacing is unity ($\lambda=1$) this energy per site is 
also the energy density, and the density is the site filling fraction.

Near half filling $\varepsilon_{PM} = -(4/\pi)^2t+U/4$, which becomes
positive when the repulsive interaction exceeds $U/t\ge 64/\pi^2\simeq
6.5$. The PM phase is then no longer the ground state.  In a state with
only one spin the antisymmetry of the wavefunction automatically
removes double occupancy and the repulsive term $Un^2/4$ in Eq. (\ref{PM})
disappears. Such a ferromagnetic (FM) state always has negative energy
$\varepsilon_{FM}\le0$ for $n\le1$ and is a candidate for the
ground state. AF, linear AF \cite{Kusko} and stripe phases are other 
competing candidates. Furthermore phase separation (PS) and mixed phases 
can occur near half filling. When $U\la
7t$ the ground state of the 2D MF Hubbard model undergoes transitions
from an AF at half filling to a mixed AF+PM phase for doping $\delta=1-n$ up to a finite
($U$ dependent) value $|\delta|\le\delta_s$where after a pure PM phase takes over.
\cite{Langmann} For larger $U$ the phase diagram is more complicated
with a pure as well as mixed FM phases between the AF and PM phases. A
finite next neighbour hopping term $t'$ makes the phase diagram
asymmetric around half filling, extends the AF phase and changes the phase diagram 
considerably.

At half filling $n=1$ the ground state is an AF.
Near half filling $0\le |\delta| \ll1$
the MF equations and AF energy can be expanded as
\bea
   \varepsilon_{AF} = -J[1+\frac{3}{2}\delta+2\delta^2+...] \,. 
\eea
The concave dependence on $\delta$ signals phase separation into a mixed
phase of AF and PM by the Maxwell construction. The
PS extends from the AF phase with density $n=1$ to a PM phase.
Dynamical
mean field theory \cite{Pruschke} find that for particle or hole
doping $\delta=1-n$, phase separation occurs between an AF at half
filling and a paramagnet at a critical filling $\delta_s$ depending on the
coupling. It has a maximum $\delta_s\simeq 0.14$ around the coupling $U\simeq
9t$ where also $T_N$ is maximal. PS is not observed in cuprates. It may be
inhibited
by long range Coulomb repulsion between electrons in cuprates
as will be discussed later.

\subsubsection{d-wave superfluidity in 2D}

d-wave or $d_{x^2-y^2}$
pairing is particular favoured on a square lattice with on-site
repulsion (which disfavours s-wave pairing) and nearest neighbour
attraction as e.g. provided by the super-exchange coupling in the
Hubbard and Heisenberg model with two spin components. 
The discussion below emphasizes
how 2D and also 3D optical lattices can resolve important issues as
the existence of phase separation, stripe phases and dSF.

Repulsive onsite interactions $U>0$ inhibit s-wave pairing
unless a longer range attraction is added such as a nearest
neighbour interaction $V\sum_{\langle ij\rangle}n_in_j$.  Super-exchange in the
Hubbard model generates a similar nearest neighbour (spin-spin)
interaction with coupling $J=4t^2/U$, which is believed to be responsible 
for high temperature superconductivity \cite{Anderson}.  
In the limit $U\to \infty$
extended s-wave pairing can occur but requires a strong nearest neighbour attraction
$V<-\pi^2t/2$ \cite{Micnas,Nozieres}, whereas d-wave pairing occurs naturally 
in 2D for even weak nearest neighbour attraction. The d-wave mean field gap
equation is (see, e.g., \cite{Micnas,Nozieres})
\bea
  1 = -\frac{V}{4} \sum_{\bf q} \frac{\eta^2_{\bf q}}{2E_{\bf q}} \,,
\eea
where now $E_{\bf q}=\sqrt{\epsilon^2_{\bf q}+\Delta_d^2\eta^2_{\bf q}}$
and $\eta_{\bf q}=2[\cos q_x-\cos q_y]$.
At low filling the d-wave gap can be calculated within mean field
to leading orders in the density \cite{gos}
\bea
   \Delta_d = \frac{t}{\sqrt{n}}
\exp\left[\frac{4}{\pi n^2}\left(\frac{t}{V}+c_0+c_1n+c_2n^2\right)\right] \,,\quad n\ll1,
\eea
The higher order corrections in density are:
$c_0=4/\pi-1\simeq0.27$, $c_1=\pi/2-1\simeq0.57$ and $c_2\simeq0.09$. 

At half filling we can calculate the d-wave gap within mean field as above
if correlations can be ignored, which requires that the on-site interaction is small.
To leading logarithmic orders the d-wave pairing gap is
\bea \label{TvH}
   \Delta_d = \frac{8}{e^2} t\exp\left[-\pi\sqrt{t/|V|}\right] \,,\quad n=1\,.
\eea
where again the singular level density leads to a gap similar to Eq. (\ref{G2D}).
It is a coincidence that the prefactor $8/e^2$ is the same as in
Eq. (\ref{G3D}). Correlations are, however,
expected to suppress the d-wave mean field gap of Eq. (\ref{TvH})
near half filling when the on-site repulsion is strong.

$T_c$ can in 3D be calculated from the
gap equation (\ref{sgap}) whereas in 2D a
Berezinskii-Kosterlitz-Thouless transition occurs \cite{Scalettar} at
a lower temperature $T_{BKT}$. At low density the transition
temperature is in 3D generally proportional to the gap with the same
prefactor $T_c/\Delta=e^{C_E}/\pi\simeq0.567$ 
for both s- and d-wave superfluidity with and without
induced interactions \cite{PS}. In 2D a mean field critical
temperature $T_{MF}>T_{BKT}$ \cite{Scalettar}, can be calculated from
the gap equation as above and we find that the same ratio applies even
near half filling where logarithmic singularities appear. This implies
that induced interactions change the mean field critical temperatures
by the same factor as the gap of Eq. (?).

\subsubsection{t-J-U model predictions}

\begin{figure}[t]
\includegraphics[scale=0.92,angle=-90]{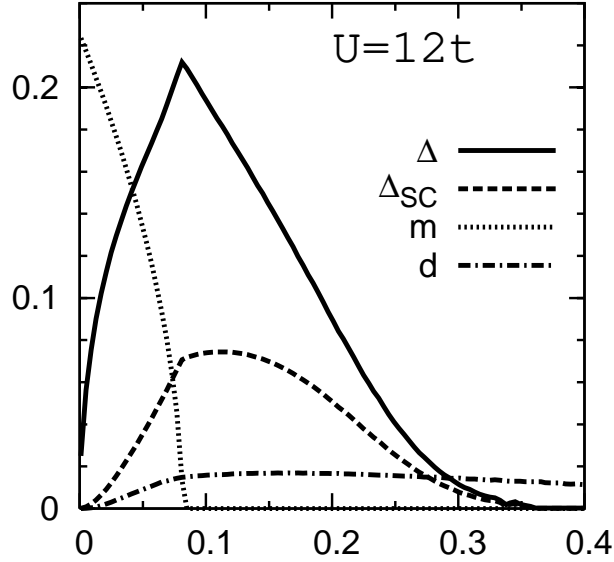}
\caption{Order parameters for dSF ($\Delta_{SC}$), AF (m) and double occupancy (d) 
vs. doping $\delta=1-n$ for 
$J/t=4t/U=1/3$ in RMFT for the 2D t-J-U model. From \cite{gos}.}
\label{Figgos}
\end{figure}

The various competing phases
can be studied in the 2D t-J-U model within the Gutzwiller projection
method and renormalized mean field theory (RMFT) \cite{gos}.
This method approximates the strong correlations and generally agrees well with
full variational Monte Carlo calculations \cite{Edegger,Laughlin}.
RMFT predicts phase separation near half filling
between an antiferromagnetic (AF) Ne\'el order and a d-wave superfluid
(dSF) phase for sufficiently strong onsite repulsion.

The t-J-U model was employed by Laughlin, Zhang and coworkers
\cite{Laughlin} to study AF, HTc and ``gossamer
superconductivity'' in cuprates and organic superconductors. Both the
Hubbard and t-J models are included in the t-J-U Hamiltonian
$H=H_U+H_t+H_s$ or
\bea \label{tJU}
 H   &=& U\sum_i \hat{n}_{i\uparrow} \hat{n}_{i\downarrow}
-t\sum_{\av{ij}\sigma} \hat{a}_{i\sigma}^\dagger \hat{a}_{j\sigma} \,+\,
  J\sum_{\av{ij}} {\bf S}_{i} {\bf S}_{j}\,,
\eea
where $\hat{a}_{i\sigma}$ is the usual Fermi 
creation operator, $\sigma=(\uparrow,\downarrow$) is the two hyperfine states 
(e.g. ($-\frac{9}{2},-\frac{7}{2}$) for $^{40}$K), 
$n_{i\sigma}=\hat{a}_{i\sigma}^\dagger \hat{a}_{i\sigma}$ the density, 
${\bf S}_i=\sum_{\sigma\sigma'}\hat{a}_{i\sigma}^\dagger\vec{\sigma}_{\sigma\sigma'}
\hat{a}_{i\sigma'}$
and $\av{ij}$ denotes nearest neighbours.  $U$ is the on-site 
repulsive interaction, $t$ the nearest-neighbour hopping
parameter and $J$ the spin-spin or super-exchange coupling. 

The t-J-U model allows for doubly occupied sites and thereby also MI
transitions. As both are observed in optical lattices the t-J-U model
is more useful as opposed to the t-J model which allows neither. 
For large $U/t$ the
t-J-U and Hubbard models reduce to the t-J model with spin-spin
coupling $J=4t^2/U$ due to virtual hopping. At finite $J$ and $U$ the
t-J-U model is to some extent double counting with respect to the
Hubbard model with $J=0$. However, when RMFT is applied the virtual
hopping and resulting spin-spin interaction vanishes 
in the Hubbard (t-U) model, i.e. $J=0$, which justifies the explicit
inclusion of the spin Hamiltonian as done in the t-J-U model.

The AF phase dominates near half filling suppressing the superfluid
order parameter. The AF order parameter is
$m=(\sqrt{3}/2)\sqrt{1-T/T_N}$ at temperatures just below the
critical Neél temperature $T_N=2J$.
The d-wave superfluid gap depends sensitively on coupling, magnetization
and double occupancy as seen in Fig. (\ref{Figgos}). It competes with
the AF order parameter and dominates slightly away from half filling
as seen in Fig. (\ref{Figgos}).

\begin{figure}[t]
\includegraphics[scale=0.8,angle=-90]{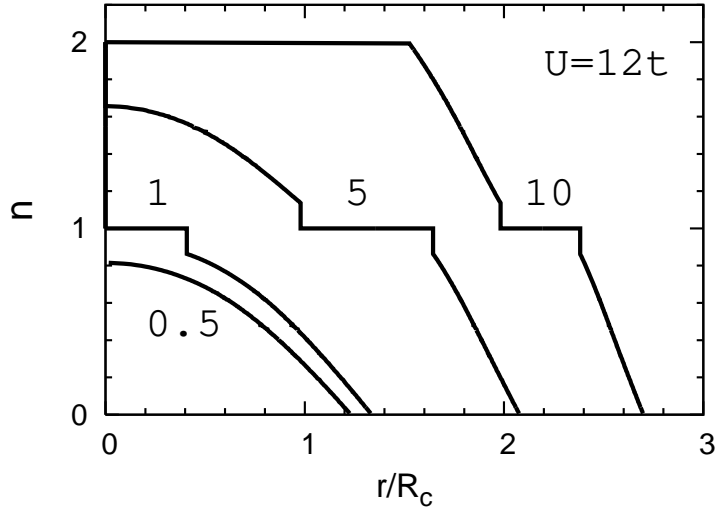}
\caption{Density distributions of Fermi atoms in a zero temperature 2D optical lattice
confined by a harmonic trap for $U=12t$ ($J=t/3$) filled with the number
of particles $N/N_c=$ 0.5, 1, 5, and 10. The density discontinuities from 
$n\simeq 1.14$ and $n\simeq0.86$ to the antiferromagnetic insulator at $n=1$ due to PS between 
dSF and AF are absent for stripe phases (see text). Thus the density distribution
changes to a {\it Capitol} form rather than the usual {\it wedding cake} 
form for Mott insulators.}
\label{Figdens}
\end{figure}

However, near half filling phase separation is found
between an AF phase at half
filling coexisting with a dSC phase at a density $|x|\la 0.12$
as determined by the Maxwell construction.  The PS is
found to terminate at coupling $J\simeq0.55t$ ($U\simeq7.3t$) where
the double occupancy undergoes a first order transion from zero to a
finite value. The PS has a curious effect on the density 
distribution in a trap as shown in Fig. (\ref{Figdens}).
The trap size is of order $R_c=\sqrt{8t/\hbar\omega}a_{osc}$, 
and the number of trapped particles
$N=2\pi\int^R_0 n(r)rdr$ is of order $N_c=\pi R_c^2$ in 2D.  

The density distributions and MI plateaus have been measured
experimentally for Bose atoms in optical lattices by, e.g.,
differentiating between singly and doubly occupied sites
\cite{Folling}. Recently Schneider et
al. \cite{Schneider} have measured column densities for Fermi atoms in
optical lattices and find evidence for incompressible Mott and band
insulator phases. Even lower temperatures are required for
observing the AF and dSF phases and the density discontinuities due to PS.

Spin and charge density waves in form of stripes are not included in
the above RMFT calculations. Stripes are observed in several cuprates
\cite{Tranquada} whereas numerical calculations are model dependent. 
Long range Coulomb frustration can explain why PS is
replaced by stripes and an AF phase at very low doping \cite{gos} as
observed in cuprates.

\subsection{High temperature superconductivity vs. optical lattices}

After Kamerlingh Onnes discovered superconductivity in 1911 it took
nearly half a century before Bardeen, Cooper and Schriefer (BCS) in
1957 understood the physics behind the manybody pairing mechanism.  In
comparison half the time has passed since Bednorz \& M\"uller in 1986
discovered high temperature superconductivity (HTc).  Although
important progress has been made on the mechanisms behind this
important phenomenon, a full understanding is still lacking.  Most
agree that HTc is d-wave and can probably be described by the one band
Hubbard model on a two-dimensional (2D) lattice with strong repulsive
on-site interaction $U\sim3t$, but no consensus has been reached on
the origins and influence of the pseudogap, charge and spin gaps,
stripes, etc.

\begin{figure}[t]
\includegraphics[scale=0.65,angle=-90]{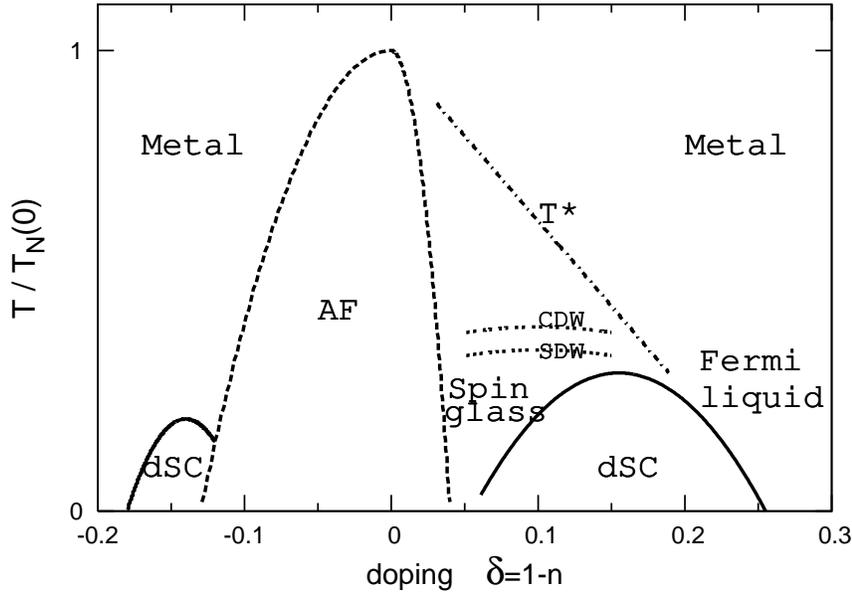}
\caption{Typical phase diagram of high temperature
superconductors.
Ne\'el temperatures ($T_N$) and critical temperatures for d-wave 
superconductivity ($T_c$) are plotted vs. particle ($\delta<0$) and
hole ($\delta>0$) doping. $T^*$ is the pseudogap (see text). }
\label{Cuprate}
\end{figure}

A sketch of a typical phase diagrams of HTc cuprates is shown
in Fig. (\ref{Cuprate}). The antiferromagnetic phase extends
around half filling and is asymmetric due to next neighbour hopping
terms $t'$. The Ne\'el temperature drops rapidly from its maximal
value $T_N(\delta=0)$ - somewhat lower than the Fermi temperature.
The maximal $T_c\la100$K are collected in the ``Uemura'' plot \cite{Uemura}.
It is typically between one and two orders of magnitude
lower than the Fermi energy which again is several orders of
magnitude larger than conventional superconductors as Nb, Sn, Al, Zn.

Optical lattices hold great promises for solving the HTc problem 
\cite{Hofstetter,Chen} since
densities, temperatures, interactions, spins, etc.  can be varied in a
controlled manner and studied in experiments, where the sub nanometer
lattice constants in solids are upscaled to micrometers in optical
lattices. One serious experimental hindrance remains, namely the extremely low
temperatures required, but we have hopes that the experimentalists can
overcome this problem rapidly as in so many other cases with cold atoms - 
and it certainly does not hinder theoretical studies.

\subsubsection{Stripes}

Stripes have been discovered in low-energy magnetic neutron scattering
in doped cuprates at incommensurate longitudinal and horizontal charge
and spin wave numbers \cite{Tranquada}, e.g. ${\bf
  Q}_c=(2\pi/\lambda)(0,\pm \delta/\delta_s)$ and ${\bf
  Q}_s=(\pi/\lambda)(1,1\pm \delta/\delta_s)$ for the longitudinal
charge and spin wave numbers respectively. Here, the average doping or
holes are all in the stripes with hole filling $\delta_s$ embedded in
an AF with filling no holes (half filled) such that the average hole
density is $\delta=1-n$.  The stripes are half filled $\delta_s=1/2$
for low average dopings $\delta\le 1/8$ but at larger doping it
increases linearly to filled stripes $\delta_s=1$ such that
$\delta/\delta_s=1/4$ remains constant.  Therefore the charge (spin)
density stripes appear with periodic distance $d$ ($2d$) depending on
doping as $d=\lambda\delta_s/\delta$.  The stripe distance decreases
with increasing doping until $\delta\ge1/8$, where after the stripes
remain at a distance $d=4\lambda$. The stripes act as anti-phase
domain walls and the spin density wave therefore has periodicity twice
the length of the charge-density wave. 
When $1/8\le\delta\le1/4$ the 2D spin orientation therefore looks like
\bea
   ...\downarrow\uparrow\downarrow\circ\uparrow\downarrow\uparrow\circ
   \downarrow\uparrow\downarrow\circ\uparrow\downarrow\uparrow\circ...\nonumber\\
   ...\uparrow\downarrow\uparrow\circ\downarrow\uparrow\downarrow\circ
   \uparrow\downarrow\uparrow\circ\downarrow\uparrow\downarrow\circ...\nonumber\\
   ...\downarrow\uparrow\downarrow\circ\uparrow\downarrow\uparrow\circ
   \downarrow\uparrow\downarrow\circ\uparrow\downarrow\uparrow\circ...\nonumber
\eea
etc., in case of vertical stripes $\circ$. They are spin-balanced
at filling $1-\delta_s$ and occur with periodicity $d=4\lambda$.

Diagonal, horizontal/vertical, checkerboard stripe solutions have been
found in a number of models. In MF models Zaanen and Gunnarson
\cite{Zaanen} found stripes of hole density $\delta_s=1$ that are vertical
or horizontal for $U/t\la 3-4$ and diagonal otherwise.  In DMRG
calculations \cite{Scalapino} stripes with hole density $\delta_s=1/2$ are
found in agreement with experiments \cite{Tranquada}.
Stripes appear at temperatures above HTc but below the pseudogap $T^*$
\cite{Berg} and there are some evidence that charge density waves
appear at slightly higher temperatures than spin density waves.

The stripe phase is a specific ordered mixed AF and PM phase and is a
continuous transition between the two pure phases as function of
density.  Similar mixed phase solutions are believed to occur in
neutron star crusts between nuclear matter and a neutron gas
and possibly also between quark and nuclear matter \cite{Ravenhall}.
In both cases Coulomb energies add complexity to the mixed phases by ordering
them into structured crystallic phases.

The various MI, AF, stripe, d- and s-wave superfluid phases have distinct
density and momentum correlation functions 
\cite{Ketterle,Kohl,Folling,Spielman,Rom,Bruun,gos}.

\subsubsection{Coulomb frustration}

Long range Coulomb interactions prevent phase separation into two bulk
phases of different charge density but may not inhibit 
the formation of localized holes, pairs and stripes. The phase diagrams
of cuprates with such Coulomb frustration and optical lattices without
may therefore be very different.
Coulomb frustration in cuprates has been discussed in connection with
stripes (see e.g. \cite{Seibold,Fradkin}). 

In the following we shall consider the stripes as
rods with charge less than the surrounding phase and calculate the
additional Coulomb energy of such structures.
Coulomb energy densities have been calculated for structures of various dimensionality
$D$ and volume filling fraction $f$ \cite{Ravenhall}
\bea \label{Ec}
   {\varepsilon}_C = \frac{2\pi}{D+2}\Delta\rho^2 R^2 
    f\left[ \frac{2}{D-2}(1-\frac{D}{2}f^{1-2/D})+f \right] \,.
\eea
Here, the charge density difference between the two phases is
$\Delta\rho = e\delta_s/\lambda^3$ for the stripes and the volume filling fraction is  
$f=\delta/\delta_s=\lambda/d$.
The dimensionality is $D=3$ for spherical droplets or bubbles,
$D=2$ for rods and tubes and $D=1$ for plate-like structures.
The diameter of the spheres, rods or the thickness of the plates is of order the
distance between layers $2R\sim  \lambda\sim 4$\AA.
The stripes are rods in a 2D plane but are embedded in a 3D
layered structure.
For $D=2$ the expression in the square bracket of Eq. (\ref{Ec}) reduces
to $[\ln(1/f)-1+f]$. The logarithm originates from the Coulomb integral
$\int^ldz/z$ along the rod length $z$, which is cutoff by other rods at
a length scale $l\sim \lambda\sqrt{f}$.
For the cuprates we furthermore reduce the Coulomb field by a dielectric
constant of order $\epsilon_D\sim5$.
The resulting Coulomb energy of stripe or rod-like structures $D=2$ is 
\bea \label{Ecc}
  \varepsilon_C \simeq \frac{\pi}{8}\frac{\delta_se^2}{\lambda^4\epsilon_D} 
   f \left[\ln(1/f)-1+f\right] \,.
\eea

Energy costs associated with the interface structures are usually
added. Such surface energies are difficult to calculate for the stripes
because their extent is only a single lattice constant. In principle
they are already included in the stripe models. We will therefore
just add the Coulomb energies given above. However, the Coulomb energies and
the energy of the systems as a whole,
may be reduced by screening and hole hopping into the AF
whereby $R$ increases but $\Delta\rho$ and $f$ are reduced.

Inserting numbers $e^2/\hbar c=1/137$, $\epsilon_D=5$, $\delta_s=1/2$,
$\lambda=4$\AA, we find that $\varepsilon_c\simeq
150$~meV$(\delta/\lambda^3)[\ln(1/f)-1+f]$.  In comparison the energy
gain by changing phase from an AF to a stripe or PM phase increases
with doping as $\sim J\delta/\lambda^3$, where $J\simeq
(0.1-0.2)t\simeq 50-100$~meV in the cuprates. The Coulomb energy of
Eq. (\ref{Ecc}) thus dominates at small doping due to the logarithmic
singularity and therefore the AF phase of density $n\la 1$ is
preferred.

The AF phase is the ground state as long as the Coulomb energy of Eq.
(\ref{Ecc}) exceeds $Jx/\lambda^3$ corresponding to doping less than  
\bea
   \delta_{AF}\simeq \delta_s\exp
   \left[-\frac{8}{\pi}\frac{J\lambda\epsilon_D}{\delta_se^2}-1
         \right] \,.
\eea
Inserting the above numbers 
we find $\delta_{AF}\simeq 0.1$ which is within range of
the observed $|\delta_{AF}|\simeq 0.03$ for hole doped and $\delta_{AF}\simeq 0.15$
for particle doped cuprates. In MF the particle-hole asymmetry arises from the
next-nearest neighbour hopping $t'\simeq -0.3t$ and leads to
an AF phase extending from half filling up to a particle doped density
$n>1$. \cite{Langmann}

In the above picture the incommensurate stripe phases at small doping
arise due to Coulomb frustration when $t'=0$. At larger doping the stripes approach
each other and will eventually affect each other. Experimentally the stripes
undergo a transition from an incommensurate to a commensurate phase
at $\delta\simeq 1/8$ corresponding to a stripe periodicity of
four lattice spacings.

 The Mermin-Wagner theorem states that a continuous
symmetry cannot be spontaneously broken at finite temperature in two and lower
dimensional systems for sufficiently short
range interactions \cite{MW}, and as a consequence there can be no phase transition at
finite temperature. In stead a Berezinskii-Kosterlitz-Thouless
transition can occur as found in the 2D Hubbard model
\cite{Scalettar}.  This is incompatible with the phase diagram of HTc
cuprates where AF, dSC and stripe phase transitions are
observed. These paradoxial transitions and the origin of the phases
are still unresolved problems which are largely avoided in the literature. Some suggest
that inter-planar couplings effectively increase the dimension above
two. Others that correlations lengths are so 
long that they effectively looks like real transitions. 
Another suggestion in line with
the Coulomb frustration discussed above is that Coulomb interactions
are responsible for the phase transitions to stripes and d-wave
superconductivity. The long range Coulomb interactions are not
comprised in the Mermin-Wagner theorem, which only applies to sufficiently short range
interactions $U(r)\propto r^{-\alpha}$ with $\alpha\ge D+2$
\cite{MW}.

\section{Summary and Outlook} %

In the last decade the physics of cold atoms has brought
important understanding of the unitarity limit, universality, the BCS-BEC
crossover, and many other properties of cold Fermi atomic systems as
describe above and elsewhere in this volume.  Presently a number of
other interesting phenomena are investigated such as the three component
system of $^6$Li and other multicomponent systems, the ferromagnetic
transition at large positive (repulsive) scattering lengths, spin
polarized systems, phases in optical lattices, etc.

In these lecture notes the main topic has been universality in crossover.
Not only the BCS-BEC crossover in uniform system but also
in the repulsive ``ferromagnetic'' crossover, in multicomponent systems, traps and lattices.
The Fermi particles have mostly been atoms but applications to neutron, nuclear and quark
matter, nuclei and electrons in solids have been made wherever possible.

In view on the important impact that the question Bertsch posed a
decade ago made, we may attempt to ask new relevant questions:
\begin{svgraybox}
 What are the universal thermodynamic functions and parameters
for positive scattering lengths and
is there a ferromagnetic phase transition as $a\to +\infty$?
\end{svgraybox}
\noindent
- in both uniform systems and traps.
One opinion is given above but a dispute is ongoing \cite{Ho,Jo}.

A most promising direction is optical lattices where we may ask:
\begin{svgraybox}
 What are the phase diagrams of the two and three dimensional Hubbard models as
realized in optical lattices?
\end{svgraybox}
\noindent
Attractive interactions will lead to a superfluid state with critical temperatures
that have been calculated in detail by Monte Carlo.
For repulsive interactions the phase diagram is not very well known in
2D or 3D even for the single band t-U or t-J model.
The antiferromagnetic phase may extend to densities near half filling before undergoing
a transition to a paramagnet
but it is competing with phase separation, 
stripe phases and a ferromagnetic phase for very strong repulsion.
In 2D it also competes with d-wave superconductivity.
 
A related question, which is particular relevant for HTc, is:
\begin{svgraybox}
Does the 2D Hubbard model (as realized in optical lattices) exhibit 
high temperature superconductivity (as seen in cuprates)?
\end{svgraybox}
\noindent
Because the electrons in HTc have charge they cannot undergo phase
separation in bulk, which in turn may be responsible for stripe
formation in cuprates in the region of densities and temperatures
where also d-wave superconductivity appears. Ultracold Fermi atoms are
neutral and can therefore undergo phase separation near half filling
as is predicted in some models, e.g. the t-J-U model with Gutzwiller
projections discussed above. 
If so, the region of densities near half filling with antiferromagnetic and d-wave
superconductivity do not exist in optical lattices.
Cuprates may also have multiband coupling and certainly
has next nearest neighbour hoppings, which cause asymmetries in the phase diagrams
around half filling for both the AF and dSC phases.
In any case, the differences between the phase diagram of the cuprates and the
Hubbard model as realized in optical lattices will reveal the
important driving mechanisms and possible additional ingredients present in
the phenomenon of HTc. 2D optical lattices may also resolve the Mermin-Wagner paradox.

In summary, we have good reasons to believe that ultracold Fermi atoms
in traps and optical lattices will in few years bring deeper
understanding of strongly correlated systems, its para-, ferro- and
antiferromagnetic phases, Mott insulators, stripes, s- and d-wave
superfluidity, and high temperature superconductivity.

%

%
%
%

\end{document}